\documentclass[draft=false,showpacs]{aa}
\usepackage{natbib,twoopt}
\usepackage{graphicx,units,multirow,subfigure,color,amsmath,amssymb,epsfig,makecell}
\usepackage{courier}
\usepackage[varg]{txfonts}
\usepackage[usenames,dvipsnames,svgnames,table]{xcolor}
\usepackage[absolute,overlay]{textpos}
\usepackage{ulem}

\newcommand{\BIG}{{\sf BIG}}
\newcommand{\SLIM}{{\sf SLIM}}
\newcommand{\QUAINT}{{\sf QUAINT}}
\newcommand{\yieldx}{{\sc yieldx}}
\newcommand{\wnew}{{\sc wnew}}
\newcommand{\galprop}{{\sc galprop}}

\newcommand{\exfor}{{\sc exfor}}
\newcommand{\usine}{{\sc usine}}
\newcommand{\minuit}{{\sc minuit}}

\newcommand{\xsS}{{\tt S01}}
\newcommand{\xsW}{{\tt W03}}
\newcommand{\xsGalxii}{{\tt Galp--opt12}}
\newcommand{\xsGalxxii}{{\tt Galp--opt22}}
\newcommand{\optxii}{{\tt OPT12}}
\newcommand{\optxiiupxxii}{{\tt OPT12up22}}
\newcommand{\optxxii}{{\tt OPT22}}
\newcommand{\nofitHE}{{\tt no-fit HE}}

\newcommand{\chidof}{\ensuremath{\chi^2/}\textrm{dof}}
\newcommand{\chimindof}{\ensuremath{\chi^2_{\rm min}}/{\rm dof}}

\newcommand{\chipernui}{\ensuremath{\chi^2_{\rm nui}/{n_{\rm nui}}}}

\newcommand{\toprule}{\hline}
\newcommand{\bottomrule}{\hline}
\newcommand{\midrule}{\hline\hline}

\bibpunct{(}{)}{;}{a}{}{,} 

\definecolor{light-gray}{gray}{0.95}
\definecolor{dark-gray}{gray}{0.4}

\usepackage[pdftex,             
            breaklinks=true,%
            colorlinks=true,%
            citecolor=blue,
            pdfauthor={Maurin et al.},%
            pdftitle={The importance of Fe fragmentation for LiBeB analyses}%
           ]{hyperref}

\begin{document}

\input epsf
\title{The importance of Fe fragmentation for LiBeB analyses}
\subtitle{Is a Li primary source needed to explain AMS-02 data?}

\author{
  D. Maurin\inst{1}\thanks{\url{david.maurin@lpsc.in2p3.fr}}
  \and E. Ferronato Bueno\inst{2}\thanks{\url{e.ferronato.bueno@rug.nl}}
  \and Y. G\'enolini\inst{3,4}\thanks{\url{yoann.genolini@lapth.cnrs.fr}}
  \and L. Derome\inst{1}
  \and M. Vecchi\inst{2}
}

\authorrunning{Maurin et al.}

\institute{
LPSC, Universit\'e Grenoble Alpes, CNRS/IN2P3, 53 avenue des Martyrs, 38026 Grenoble, France
\and Kapteyn Astronomical Institute, University of Groningen, Landleven 12, 9747 AD Groningen, The Netherlands
\and LAPTh, Universit\'e Savoie Mont Blanc \& CNRS, Chemin de Bellevue, 74941 Annecy Cedex, France
\and Niels Bohr International Academy \& Discovery Center, Niels Bohr Institute, University of Copenhagen, Blegdamsvej 17, DK-2100 Copenhagen, Denmark
}

\date{Received / Accepted}

\abstract
{High-precision data from AMS-02 on Li, Be, and B provide the best constraints on Galactic cosmic-ray transport parameters.}
{We re-evaluate the impact of Fe fragmentation on the Li, Be, and B modelling. We discuss the consequences on the transport parameter determination and reassess whether a primary source of Li is needed to match AMS-02 data.}
{We renormalised several cross-section parametrisations to existing data for the most important reactions producing Li, Be, and B. We used the \usine{} code with these new cross-section sets to re-analyse Li/C, Be/C, and B/C AMS-02 data.}
{We built three equally plausible cross-section sets. Compared to the initial cross-section sets, they lead to an average enhanced production of Li ($\sim20-50\%$) and Be ($\sim5-15\%$), while leaving the B flux mostly unchanged. In particular, Fe fragmentation is found to contribute to up to 10\% of the Li and Be fluxes. Used in the combined analysis of AMS-02 Li/C, Be/C, and B/C data, the fit is significantly improved, with an enhanced diffusion coefficient ($\sim 20\%)$. The three updated cross-section sets are found to either slightly undershoot or overshoot the Li/C and B/C ratios: this strongly disfavours evidence for a primary source of Li in cosmic rays. We stress that isotopic cosmic-ray ratios of Li (and to a lesser extent Be), soon to be released by AMS-02, are also impacted by the use of these updated sets.}
{Almost no nuclear data exist for the production of Li and B isotopes from Ne, Mg, Si, and Fe, whereas these reactions are estimated to account for $\sim 20\%$ of the total production. New nuclear measurements would be appreciated and help to better exploit the high-precision AMS-02 cosmic-ray data.}

\maketitle

\section{Introduction}

Galactic cosmic-ray (GCR) Li, Be, and B (hereafter LiBeB for short) isotopes are present in minute amount in the Solar System, but are seen in excess in cosmic-ray (CR) data \citep[e.g.][]{2018ARNPS..68..377T}. These CR species are denoted secondary species, as they are generally assumed to be produced only by a nuclear interaction of heavier CR species on the interstellar medium (ISM). The dominant channels for this production are the direct production of LiBeB from C and O CR fluxes \citep[e.g.][]{2018PhRvC..98c4611G}. The latter fluxes are among the most abundant CR species, and are of {\rm primary} origin, that is they result almost solely from the diffusive shock wave acceleration of the ISM material.
The study of secondary species, or secondary-to-primary ratios, plays a central role in CR physics because they calibrate the transport parameters in the Galaxy. The latter can give insight into the micro-physics of transport in the turbulent medium \citep[e.g.][]{2019PhRvD..99l3028G}, but it is also a central ingredient for many related studies (e.g. electron and positron spectra, $\gamma$-ray diffuse emissions, indirect searches for dark matter in CRs).

The AMS-02 experiment on-board the International Space Station has collected an unprecedented number of CRs \citep{2021PhR...894....1A}; the data published by the collaboration reach the few percent level of precision. The LiBeB data, which display a spectral break at $\sim 200$~GV \citep{2018PhRvL.120b1101A}, have been used by several authors to study the transport of GCRs \citep{2017PhRvL.119x1101G,2019PhRvD..99l3028G,2019PhRvD..99j3023E,2020A&A...639A.131W,2020ApJS..250...27B,2020ApJ...889..167B,2020JCAP...11..027Y,2021JCAP...03..099D}. The above studies assume no extra source of LiBeB, but it remains possible to have a small amount of secondary production inside the acceleration site \citep{2021PhRvD.104j3029M,2021ApJ...917...61K}, or even to have a primary source of Li \citep{2018PhRvL.120d1103K} from nova explosions \citep{2015Natur.518..307H}.

All model calculations rely on a network of CR fragmentation reactions. Uncertainties on these reactions range from $10\%$ to $20\%$, and are a limiting factor to take full advantage of the high-precision CR data \citep{2018PhRvC..98c4611G}. For this reason, possible excesses or mismatches (between the model and the data) must be robustly checked against nuclear uncertainties (among others). A new methodology to account for and propagate these uncertainties was proposed in \citet{2019A&A...627A.158D}. The main idea was to start from a given production nuclear dataset and, while fitting LiBeB data, to allow the most relevant reactions to vary around their central values; penalties prevent cross sections from wandering far away from their expected values and uncertainties. We successfully used this approach in \citet{2020A&A...639A.131W} to show that models with a pure secondary LiBeB component give an excellent fit to the AMS-02 data, provided that a few cross-section values were varied by $\sim15\%$; similar but independent studies reached the same conclusions \citep{2021JCAP...03..099D,2021JCAP...07..010D,2021PhRvD.103j3016K}. Conversely, \citet{2020ApJ...889..167B} recently argued for a hint of primary Li in CRs, though they do not exclude an overall bias in the Li production cross sections.

At a first glance, there is not much to do to improve the main LiBeB production channels. Indeed,  comprehensive inspections of the nuclear cross-section parametrisations were carried out by several independent teams twenty years ago \citep{1998ApJ...501..911S,2003ApJS..144..153W,2003ICRC....4.1917M,2003ICRC....4.1969M}, and were also re-inspected in more recent studies \citep{2018JCAP...01..055R,2018JCAP...07..006E}. However, a closer look shows that \citet{2018JCAP...07..006E} only focussed on progenitors up to O, while \citet{2018JCAP...01..055R} did so on those up to Si. This means that `recent' nuclear data, for reactions involving nuclei heavier than Si (up to Ni), may have been skipped in this re-evaluation. The most comprehensive set of parametrisations, taking advantage of all nuclear data up to Ni, are those of Webber \citep{2003ApJS..144..153W} and of the \galprop{} team\footnote{\url{https://galprop.stanford.edu/}}. These parametrisations were, for instance, extensively used in \citet{2018PhRvC..98c4611G} to rank the most important production channels for LiBeB; in these ranking, Fe (as a progenitor) only appears at the percent level. A closer inspection of the nuclear physics literature indeed indicates that important data, related to the fragmentation of Fe into LiBeB, only became available after these parametrisations were established; this means that Fe fragmentation into LiBeB must be re-evaluated.

This paper is organised as follows: in Sect.~\ref{sec:modelling}, we recall the setup used for the propagation and the motivation for our renewed interest in Fe fragmentation for LiBeB calculations. In Sect.~\ref{sec:XS}, we gather and use available nuclear data to update cross-section parametrisations of the most important progenitors, highlighting the enhanced production of Li (with regard to Be and B) from Fe. In Sect.~\ref{sec:xs_ranking}, we show a ranking of the most important progenitors of LiBeB, providing a complementary view of the rankings proposed in \citet{2018PhRvC..98c4611G}. In Sect.~\ref{sec:LiBeB_reanalysis}, we discuss the impact updated cross sections have on the derived transport parameters. In particular, we repeat and update our recent analysis of AMS-02 LiBeB data \citep{2020A&A...639A.131W}, in order to address the interpretation of a primary source of Li in CRs. We also illustrate how isotopic ratios of Li, Be, and B fluxes (e.g. $^6$Li/Li and $^7$Li/Li) are sensitive to the uncertainties on the production cross sections. We then conclude in Sect.~\ref{sec:conclusions}.

Additional comparisons and checks can be found in the appendices. App.~\ref{app:impact_per_progenitor} identifies the progenitors whose updated cross sections make up most of the observed changes in the Li/C, Be/C, and B/C ratios (hereafter LiBeB/C for short). App.~\ref{app:impact_AMS} shows that the use of the new AMS-02 F, Ne, Na, Mg, Al, Si, and Fe data only marginally impact the LiBeB/C calculations, compared to our previous analysis \citep{2020A&A...639A.131W} in which only AMS-02 C, N, and O data were available. App.~\ref{app:update_transport} shows the constraints set by the updated cross sections on various combinations of the transport parameters (only the simplest transport configuration is discussed in the main text).


\section{Propagation setup and ingredients}
\label{sec:modelling}

\subsection{CR flux calculation}
\label{sec:setup}

The quasi-isotropic flux $\psi^k(E)$ of a CR ion is related to its differential density $N^k(E)$ by $\psi^k=vN^k/(4\pi)$, with $N^k$ calculated from the transport equation \citep{1990acr..book.....B}. In this study, we consider a 1D geometry \citep[e.g.][]{2010A&A...516A..67M}, with a thin plan (half-thickness $h=100$~pc) in which the sources and the gas lie, hence where energy losses and nuclear reactions occur. We assume that CRs diffuse isotropically in a thick halo of half-size $L$, with a diffusion coefficient $K(R)$; without loss of generality for our analysis, we set $L=5$~kpc \citep{2020A&A...639A..74W,2020A&A...639A.131W}. Further assuming steady-state, the transport equation in this geometry becomes a second order differential equation along the vertical spatial coordinate and on energy.
To highlight the role of nuclear interactions, which are the prime focus of this study, we write below the (simplified) equation for our model (energy loss and gain terms, decay terms, and convective transport omitted):
\begin{eqnarray}
  -K \frac{d^2\!N^k}{dz^2}
 \!=\!  2h\,\delta(z) \left[Q^k\! +\!\!\!\sum_{i\,\in\rm ISM}\!\! n_i\left(\sum_s v^s\,\!\sigma^{s+i\to k}_{\rm prod}N^s\!-v^k\,\!\sigma^{k+i}_{\rm inel}N^k\right)\right].\!\!
\label{eq:1D}
\end{eqnarray}
The left-hand side term corresponds to diffusion, while the right-hand side has a primary source term, $Q^k(E)$, and two terms associated with nuclear interactions, written generically as $\sum_i n_iv\,\sigma(E)$. We use $n_{\rm ISM}=1~{\rm cm}^3$ ($90\%$ H and $10\%$ He in number) and $\sigma$ is a nuclear interaction, either corresponding to a net loss (destruction cross section $\sigma_{\rm inel}$) or to a secondary source term (production cross section $\sigma_{\rm prod}$, straight-ahead approximation assumed); the latter must sum over all possible CR progenitors $s$ (heavier than the produced fragments of interest $k$).
For CR elements below Ni, the above set of equations couples $\sim 100$ species, which involves a nuclear network of $\gtrsim1000$ reactions. The above coupled set of equations can be solved starting from the heaviest species---which is assumed to be a primary---and solving down to the lightest species. This ensures that all secondary contributions (from heavier species) are accounted for at each step $k$.

All the analyses presented below rely on the \usine{} package\footnote{\url{https://lpsc.in2p3.fr/usine}} \citep{2020CoPhC.24706942M}, and we stress that the analyses below solve the full transport equation, that is with energy gains and losses, decay terms, and possibly a constant convection term in the 1D geometry \citep{2019PhRvD..99l3028G}.\footnote{The same model and approach was used in several of our previous analyses of AMS-02 B/C and LiBeB/C nuclei \citep{2017PhRvL.119x1101G,2019PhRvD..99l3028G,2020A&A...639A.131W}, and we refer readers to these publications for more details.} Inelastic cross-sections are an important ingredient of the calculation, but their uncertainties are subdominant in the total error budget \citep[e.g.][]{2019PhRvD..99l3028G}, so we do not discuss them in this paper; the parametrisation of \citet{1997lrc..reptQ....T,1999STIN...0004259T} are used.

\subsection{Transport parameters}
\label{sec:ingredients}

Even in the simple propagation framework used, transport parameters come in different flavours. For instance, \citet{2019PhRvD..99l3028G} explored several configurations, with or without convection and re-acceleration, and different parametrisations of the diffusion coefficient $K(R)$. It was found that several of them were able to give a very good fit to LiBeB data \citep{2020A&A...639A.131W}. To ease the discussion and without loss of generality, most of our calculations will be carried out in the so-called \SLIM{} configuration. The latter consists of pure diffusion (no convection, no reacceleration) with a break at both low- \citep{2019PhRvD..99l3028G,2019PhRvD.100d3007V,2020A&A...639A.131W} and high-rigidity \citep{2017PhRvL.119x1101G,2019PhRvD..99l3028G,2018JCAP...01..055R,2020JCAP...01..036N} in the diffusion coefficient:
\begin{equation}
  \label{eq:def_K}
  K(R) = {\beta} K_{0} \;
  {\left\{ 1 + \left( \frac{R_l}{R} \right)^{\frac{\delta-\delta_l}{s_l}} \right\}^{s_l}}
  {\left\{  \frac{R}{1\,{\rm GV}} \right\}^\delta}\,
  {\left\{  1 + \left( \frac{R}{R_h} \right)^{\frac{\delta-\delta_h}{s_h}}
    \right\}^{-s_h}}.
\end{equation}

The high-rigidity break parameters $R_h$, $\delta_h$, and $s_h$ are fixed to the values found in \citet{2019PhRvD..99l3028G}; we also enforce $s_l=0.04$. We stress that for the rest of this section, the other parameters are fixed to $\delta_l=-0.74$, $R_l$=4.53~GV, and $\delta=0.51$, $K_0=0.0389$~kpc$^2$~Myr$^{-1}$ \citep{2020A&A...639A.131W}. We later see in Sect.~\ref{sec:LiBeB_reanalysis} how these parameters are impacted when re-analysing the LiBeB data with updated cross-section values. We also discuss how these results change with alternative convection and reacceleration models in App.~\ref{app:update_transport}.

\subsection{Nuclear cross sections: relevance of Fe for LiBeB?}
\label{sec:impact}

Production cross sections are the backbone of CR studies, and several parametrisations are publicly available from the \galprop{} or \usine{} codes: (i) \xsW{} \citep{2003ApJS..144..153W} relies on an unreleased version of the \wnew{} code \citep{1998ApJ...508..940W,1998ApJ...508..949W,1998PhRvC..58.3539W}, which is a semi-empirical parametrisation fitted on existing data; (ii) \xsS{} (A. Soutoul, private communication) is very similar to \xsW{} but with different parameters; (iii) {\tt Galp} \citep{2001ICRC....5.1836M,2003ICRC....4.1969M} is based on systematic fits on existing nuclear data, where \xsGalxii{} and \xsGalxxii{} pre-fit parametrisations come respectively from the 1998 version of the \wnew{} code or from the 2000 version of the \yieldx{} code (semi-analytical formulae fitted on data, \citealt{1998ApJ...501..911S,1998ApJ...501..920T}). More details on all these parametrisations can be found in \citet{2018PhRvC..98c4611G}.

Many LiBeB/C recent studies considered CR progenitors up to Si at most \citep{2018JCAP...01..055R,2018JCAP...07..006E,2019PhRvD..99l3028G}. Actually, most studies addressing a possible mismatch in the Li production considered parents up to Si \citep{2020A&A...639A.131W,2020ApJ...889..167B,2021JCAP...07..010D,2021PhRvD.103j3016K}, and those considering progenitors up to Fe \citep{2020ApJS..250...27B,2020ApJ...889..167B} relied on the \xsGalxii{} production cross sections.
\begin{figure}[t]
  \includegraphics[width=\columnwidth]{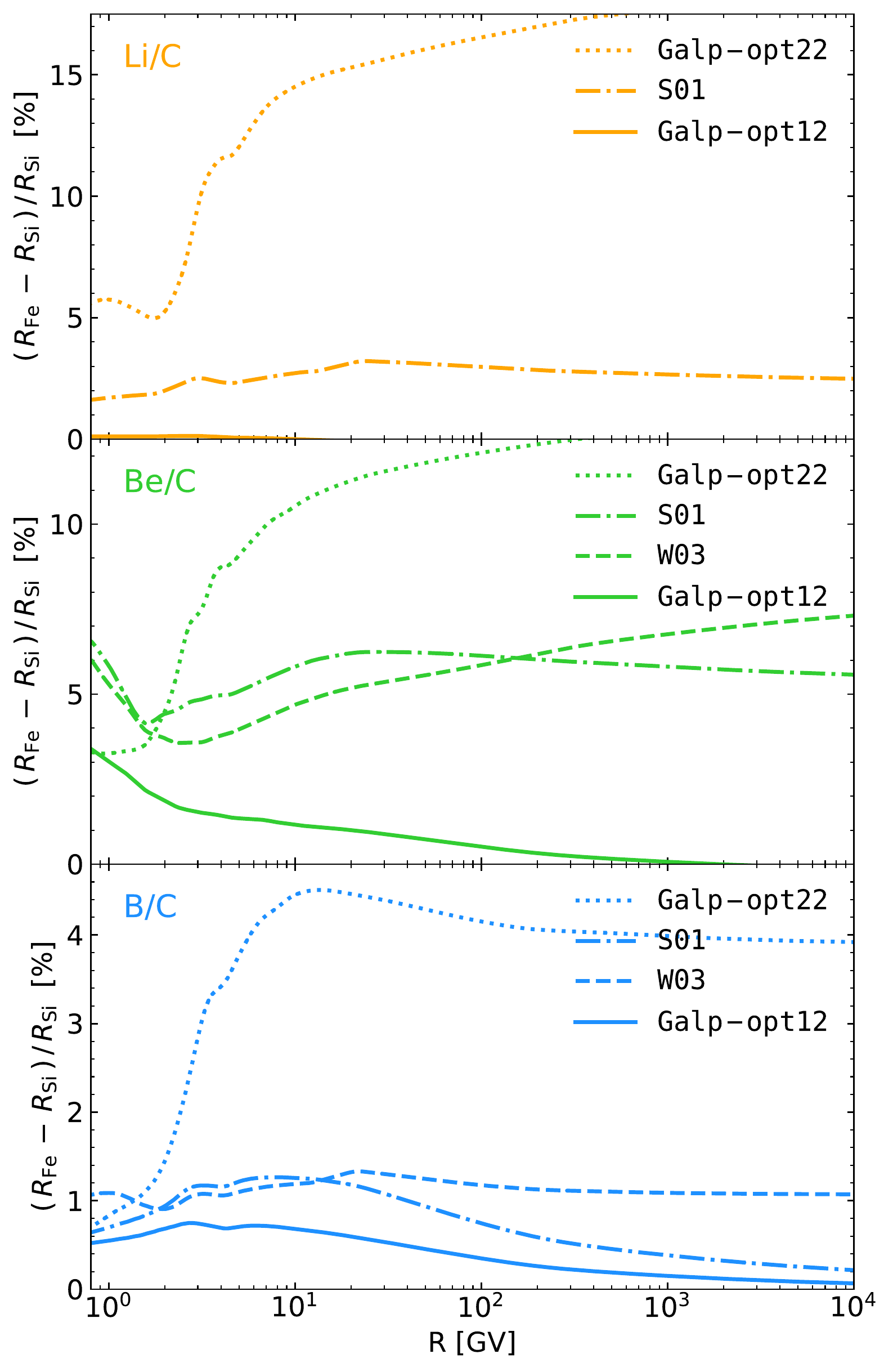}
  \caption{Relative difference on Li/C (top), Be/C (middle), and B/C (bottom) between $R_{\rm Fe}$ (progenitors up to Fe, $Z=26$) and $R_{\rm Si}$ (progenitors up to Si, $Z=14$) as a function of rigidity. The comparisons are carried out for four cross-section parametrisations available in the literature: \xsGalxxii{} (dotted lines), \xsS{} (dash-dotted lines), \xsW{} (dotted lines), and \xsGalxii{} (solid lines). The enhanced production observed with \xsGalxxii{} motivates the re-analysis of the LiBeB production cross sections.
  \label{fig:impact_heaviest}}
\end{figure}
To highlight the importance of these different settings (cross-section set and heaviest CR parent considered), we show in Fig.~\ref{fig:impact_heaviest} the relative difference between calculations of Li/C, Be/C, and B/C ratios (from top to bottom) performed with progenitors up to Fe, compared to a reference calculation with progenitors only up to Si. The various line styles and colours encode the use of the four cross-sections parametrisations discussed above, that is \xsW{}, \xsS{}, \xsGalxii{}, and \xsGalxxii{}.
The solid black lines (i.e. \xsGalxii{}) show that the choice of the heaviest progenitor (Fe instead of Si) has a marginal impact ($\lesssim3\%$) on all ratios, in line with checks that calculations up to Si are accurate enough to interpret LiBeB/C data \citep{2001ApJ...555..585M,2019PhRvD..99l3028G}. The use of \xsS{} (dash-dotted blue) and \xsW{} (dashed green) parametrisations lead to similar conclusions, though with a $\sim 5\%$ increase of Be/B (middle panel); \xsW{} does not provide production cross sections for Li. Only with \xsGalxxii{} (grey-dashed lines) do we see a significant and differential increase for Li/C ($\sim 15\%$), Be/C ($\sim 10\%$), and B/C ($\sim 5\%$). A close inspection shows that large differences exist between specific reactions for these different parametrisations, in particular for the fragmentation of $^{56}$Fe into LiBeB isotopes.

This discrepancy, corresponding to an increased production of light nuclei in Fe fragmentation, might explain the mismatch seen in several analyses of Li CR data \citep{2020A&A...639A.131W,2020ApJ...889..167B,2021JCAP...07..010D,2021PhRvD.103j3016K}. This clearly motivates a re-investigation of the nuclear data to see which of the above parametrisations is favoured by recent nuclear data.

\section{Updating the LiBeB production cross sections}
\label{sec:XS}

In this section, we review the wealth of nuclear data relevant for our study, briefly mentioning the different measurement techniques (Sect.~\ref{sec:xs_data}). We then detail a procedure to update the cross-section parametrisations, highlighting the most impacted reactions (Sect.~\ref{sec:xs_rescaling}) and bracketing by how much Fe fragmentation impacts the LiBeB/C calculations (Sect.~\ref{sec:xs_impact}).
In the rest of the paper, the original cross-section sets considered are the \xsGalxii{} and \xsGalxxii{} parametrisations, and to avoid confusion, the new (updated) cross-section sets are denoted \optxii{} and \optxxii{} respectively (a hybrid third set, \optxiiupxxii{}, is also considered).

\subsection{Nuclear data}
\label{sec:xs_data}
Most of the nuclear cross sections relevant to CR studies are from the 1980s and 1990s, when it was realised that these inputs were essential to interpret elemental CR measurements. Back then, CR data were gaining in precision over a wider energy range, a situation quite similar to the one observed nowadays (with the advent of AMS-02, DAMPE, CALET, and PAMELA instruments). On the one hand, a wealth of nuclear data was provided by the \citet{1990ICRC....3..428T}, who studied isotopic production cross sections through {\it direct reactions}, that is by measuring secondaries produced by ion beams (mostly $Z<26$) colliding liquid hydrogen or carbon and methylene CH$_2$ targets (and using a ${\rm CH}_2-{\rm C}$ subtraction technique). This technique was notably used by Bill Webber and his colleagues, in the energy range $\sim$400--800 MeV/n \citep{1982ApJ...260..894W,1988PhRvC..37.1490F,1990PhRvC..41..520W,1990PhRvC..41..533W,1990PhRvC..41..547W,1990PhRvC..41..566W,1998PhRvC..58.3539W,1998ApJ...508..940W,1998ApJ...508..949W,1996PhRvC..53..347K,1997ApJ...479..504C,1997PhRvC..56.1536C,1997PhRvC..56..398K}. On the other hand, Michel and Leya's group \citep{1989Ana...114..287M,1995NIMPB.103..183M,1997NIMPB.129..153M,1990NIMPB..52..588D,1993NIMPB..82....9B,1996NIMPB.114...91S,1998NIMPB.145..449L,2005NIMPB.229....1L,2006NIMPA.562..760L,2008NIMPB.266....2A} as well as Sisterson's group \citep{1992LPI....23.1305S,1994NIMPB..92..510S,1997NIMPB.123..324S,2000LPI....31.1432S,1998LPI....29.1234S,2002NIMPB.196..239K,2006NIMPB.251....1S} relied on {\it indirect reactions}, where a target of heavy elements is irradiated by a high-energy proton beam. With this technique, the cross sections are determined by $\gamma$-spectrometry whenever radioactive elements are produced, and after chemical processing by mass spectroscopy for the long-lived and stable isotopes. Although the target materials are generally not composed of pure isotopes, this technique gives a good approximation of the isotopic production cross sections from the dominant isotope in nature. We stress that this second technique does not directly provide the required cross sections, but must reconstruct the desired ones via measurements at different times (to account for the decay chain of short-lived nuclei)---see, for instance, \citet{2008PhRvC..78c4615T} and references therein for the steps to derive individual cross sections from cumulative ones.

The data collected for this work come from these two different techniques. Our starting point is the \galprop{} cross-section database, assembled in the file {\tt isotope\_cs.dat} of the corresponding code\footnote{\url{https://galprop.stanford.edu/}}). We then performed extensive checks of the references therein---and also a few others pointed out in \cite{2018JCAP...01..055R}---along with systematic extractions from the \exfor{} database\footnote{\url{https://www.nndc.bnl.gov/exfor/exfor.htm}} \citep{2014NDS...120..272O}. The reactions under scrutiny are those involving the most important progenitors of Li, Be, and B isotopes and ghosts (see below), that is (i) B, C, N, O, Ne, Mg, and Si, as highlighted in \citet{2018PhRvC..98c4611G}; (ii) specific production of some light isotopes from Li and Be progenitors (for completeness of the $Z\leq6$ progenitors), and (iii) Fe which is the main novelty of this study. Indeed, new relevant fragmentation data from $^{56}$Fe
\citep{2004PhRvC..70e4607N,2006NIMPA.562..729H,2007PhRvC..75d4603V,2008PhRvC..78c4615T,2011PAN....74..523T} were published seemingly after the \galprop{} database was last updated. Finally, our cross-checks lead us to replace some of the \citet{1984ADNDT..31..359R} compiled data (used in the \galprop{} file) by the original ones, and a few were even removed (because they corresponded to cumulative cross sections, i.e. with the contributions of short lives nuclei).

The updated dataset, for the specific channels leading to LiBeB, is more complete than the one used in \cite{2018PhRvC..98c4611G}. The data gathered are shown in Figs.~\ref{fig:xs_data_vs_model_1}, \ref{fig:xs_data_vs_model_2}, and \ref{fig:xs_data_vs_model_3}. Despite these additional datasets, the reader can witness the scarcity of the data at energies relevant for GCRs studies (above GeV/n) and even the absence of data for some important channels. Recent reconsideration of this issue \citep{2018PhRvC..98c4611G} has already triggered a promising new experimental programme \citep{2019arXiv190907136U} to improve the precision of these nuclear data and fill some of the gaps in existing measurements.

\subsection{From data to updated parametrisations}
\label{sec:xs_rescaling}

Providing updated cross sections and the procedure to do so depends on three factors: the energy range we are interested in, the energy coverage of nuclear cross-section data, and the underlying model assumed for the production cross sections.
The relevant energy range for AMS-02 data is a few GV to a few TeV/n, so that roughly, production cross sections of interest are in the GeV/n to TeV/n range. However, beside a few data points at 300~GeV/n \citep{1975PhLB...57..186R}, and awaiting the promising new data at 13.5~GeV/n from the NA61/SHINE collaboration \citep{2019arXiv190907136U,2022icrc.confE.102A}, the bulk of nuclear data are below a few GeV/n. Furthermore, many reactions of interest for CRs are still lacking measurements or only have a few data points. In this context, very few reactions have a full coverage in energy, that is from tens of MeV/n (reaction threshold) up to multi-GeV/n, where it is generally assumed that the cross-sections become constant with energy. For this reason, it is necessary to have an underlying model to provide the cross sections for the missing reactions, or at least to provide an energy dependence to be normalised on the few existing data. This is in spirit close to the approach followed by the \galprop{} team.

Of course, other approaches exist to improve, extend, and validate isotopic production cross sections \citep[e.g.][]{2013PhRvC..87a4606B,2021ChPhC..45h4107C,2021PhRvC.103d4606T,2021NIMPB.502..118W,2021arXiv210901388D}. These works are more focussed on the microphysics. They also have a broader scope, in order to tackle a much larger variety of reactions and situations. However, they are not necessarily optimised for CR physics. For the time being, there is not a key advantage to using these other approaches, thus we rely here on updates and rescaling of the \xsGalxii{} and \xsGalxxii{} parametrisations.

\subsubsection{Procedure}
\label{sec:procedure}
After several trials and errors, we came up with a quite flexible procedure to update the production cross sections on hydrogen. This procedure proves to be quite successful in matching the data, in order to handle (i) the variety of situations for any given reaction (completeness or scarcity of data, discrepant data), (ii) the several energy regimes of the production process (steep rise at threshold, possible bumps and deeps, and asymptotically constant behaviour at high energy), and (iii) the fact that neither a simple rescaling of the \xsGalxii{} and \xsGalxxii{} cross sections, nor a generic (and at the same time specific) enough formula can be satisfactorily used to fit all reactions. For any given reaction, this procedure goes in four steps detailed below.

\paragraph{Step 1: rescaling over the whole energy range.} To do so, (i) we calculate the ratio $r=$~data/model for the associated nuclear data; (ii) from these $r$, we form a spline ${\cal S}_r$ that is used as an interpolation function (constant if used for extrapolation); (iii) we multiply the original model by ${\cal S}_r$. A logarithmic smoothing kernel (0.15 decade in energy) is then applied to ensure that the cross sections cannot have large variations on small energy scales\footnote{This smoothing cures situations of nearby discrepant data points, while its main drawback is to slightly shift the threshold energy of the reaction to higher energy if there are no data; otherwise see Step~2.}.

\paragraph{Step 2: dedicated fit for the low-energy rise.} Near the reaction threshold, in order to reproduce the quick rise and possible bump, if enough data are available (typically $>6$ data points), we directly fit a second order polynomial in $\log-\log$ scale enforcing convexity. This fit, that accounts for data uncertainties, replaces Step~1's result in the low-energy regime.

\paragraph{Step 3: high-energy behaviour.} Above 1.5 GeV/n, we consider two alternatives. We either keep the possible energy-dependant behaviour obtained at Step~1 (referred to as \nofitHE{} later on), or enforce a constant value based on a fit (accounting for data uncertainties) on existing data at these energies\footnote{We chose 1.5 GeV/n in order to have one or multiple points to fit a constant, but this is far from being an ideal choice, in a region where some energy dependence might remain---see the prediction of the \wnew{} code in Fig.~5 of \citet{2003ApJS..144..153W} as an illustration. In any case, the residual dependence of the models is generally within the nuclear data uncertainties (see the same figure in \citealt{2003ApJS..144..153W}). Waiting for more data in this energy range, using 1.5~GeV/n or a higher value (e.g. 2~GeV/n) would change a few cross-section values here and there, but we checked that it has a minor impact on the global calculation and all our conclusions.}. The latter configuration is taken as default, used throughout our paper, unless explicitly stated otherwise.

\begin{figure*}[h!]
   \centering
   \includegraphics[trim={0 43 0 6},clip,width=0.81\textwidth]{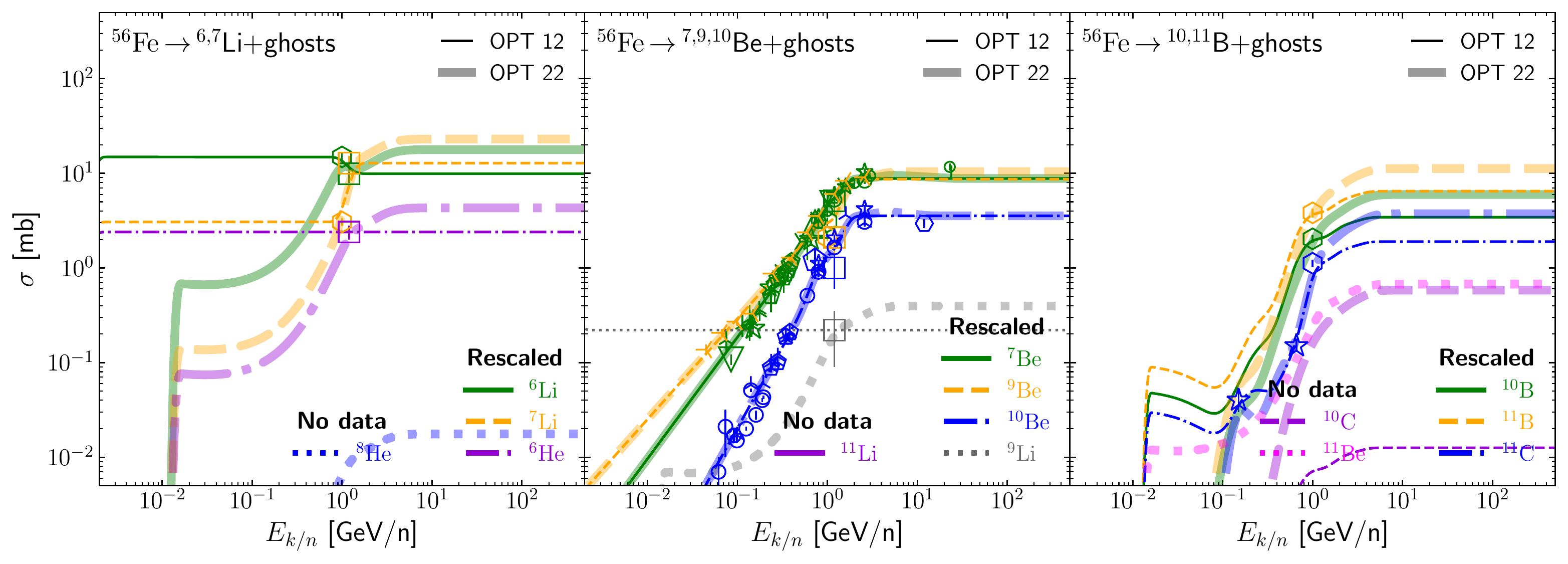}
   \includegraphics[trim={0 43 0 6},clip,width=0.81\textwidth]{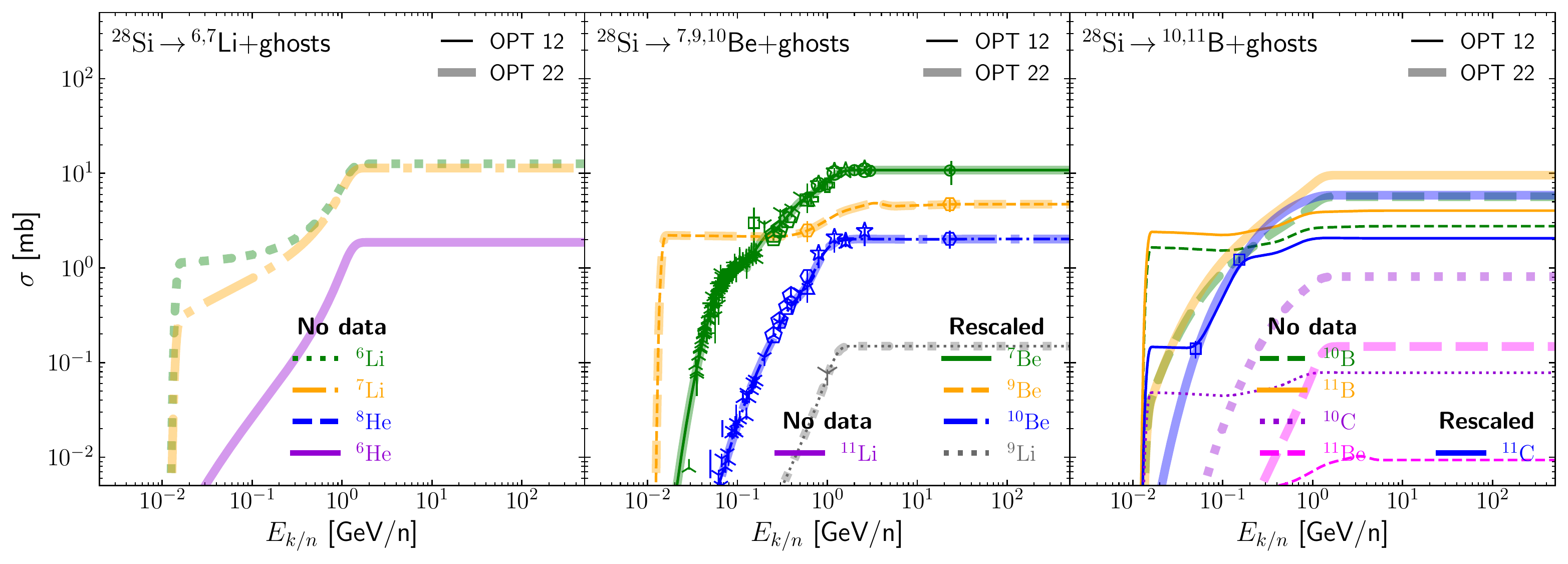}
   \includegraphics[trim={0 43 0 6},clip,width=0.81\textwidth]{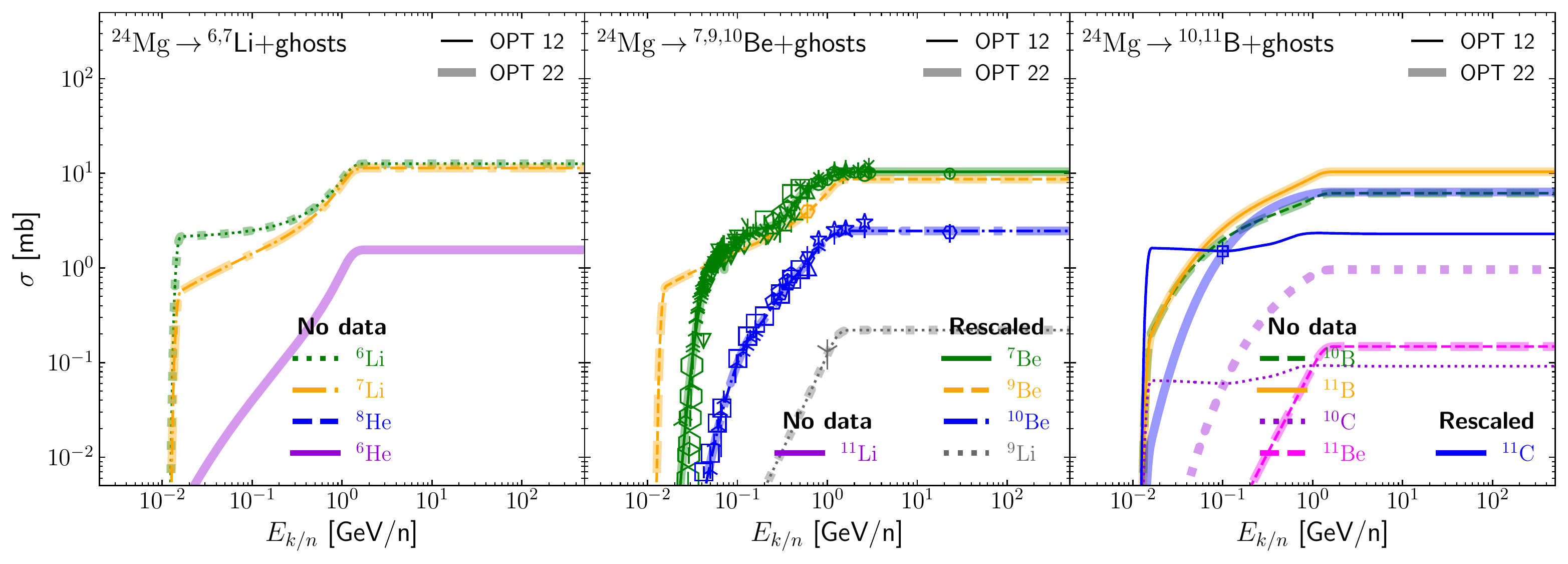}
   \includegraphics[trim={0 43 0 6},clip,width=0.81\textwidth]{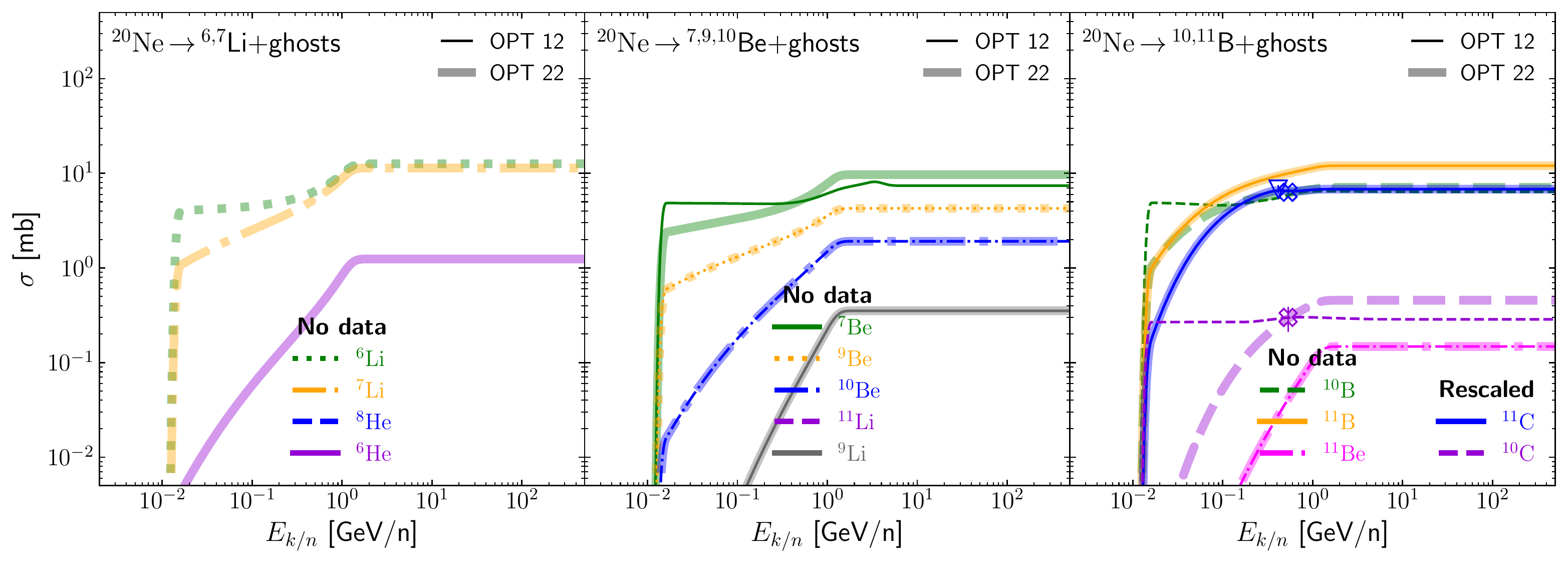}
   \includegraphics[trim={0 0  0 6},clip,width=0.81\textwidth]{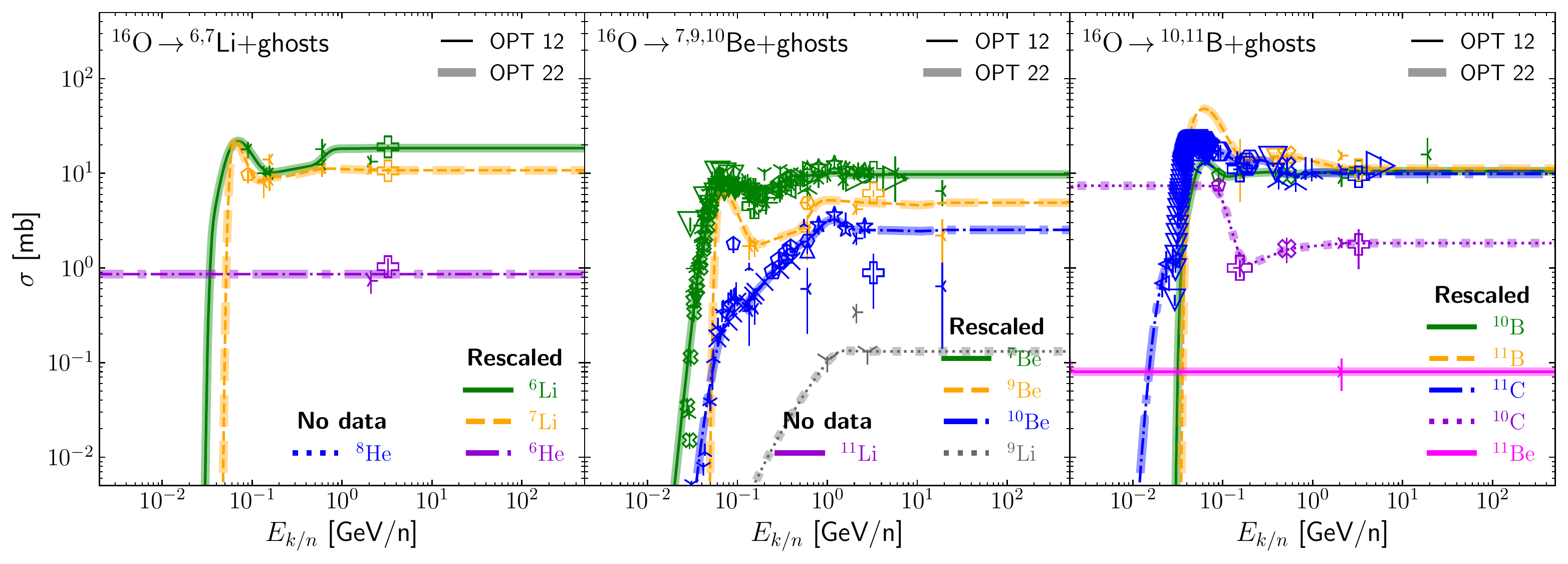}
   \caption{Comparison of models and nuclear data for the fragmentation cross sections of $^{56}$Fe, $^{28}$Si, $^{24}$Mg, $^{20}$Ne, and $^{16}$O into Li (left), Be (centre), and B (right) isotopes and associated ghosts. The references for the data points are gathered in Fig.~\ref{fig:xs_data_vs_model_leg}.
   \label{fig:xs_data_vs_model_1}}
\end{figure*}
\begin{figure*}[h!]
   \centering
   \includegraphics[trim={0 43 0 6},clip,width=0.81\textwidth]{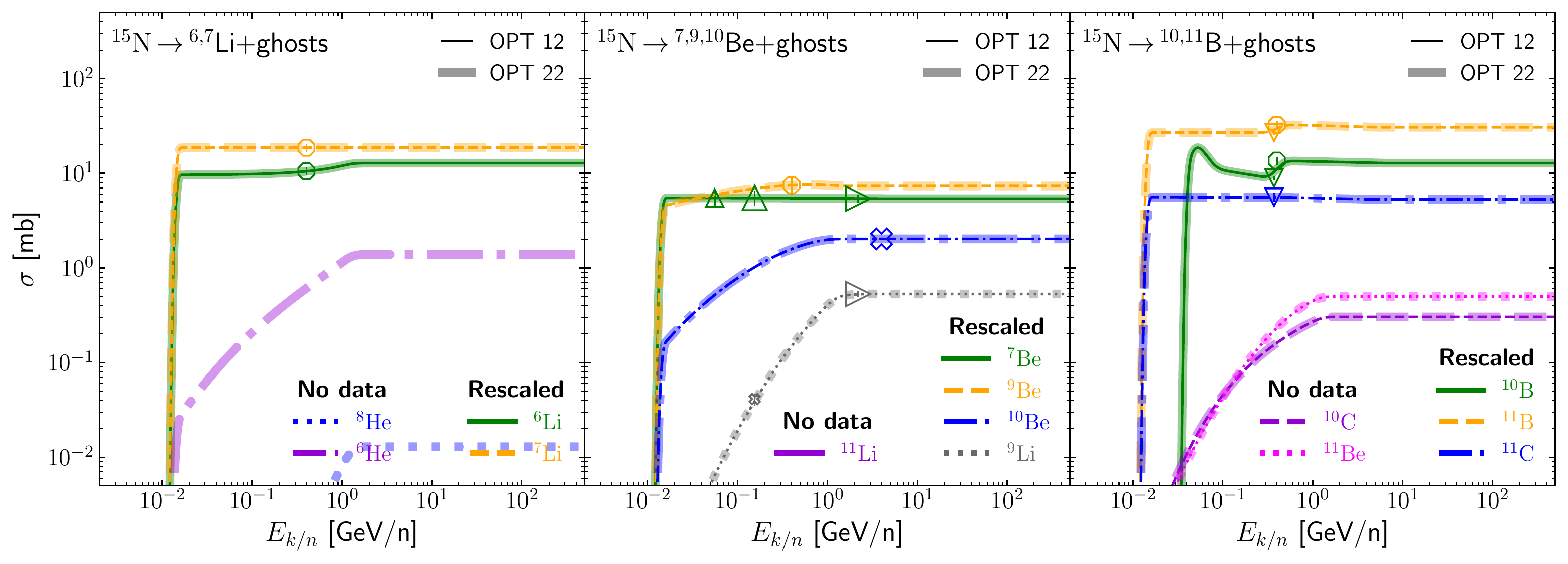}
   \includegraphics[trim={0 43 0 6},clip,width=0.81\textwidth]{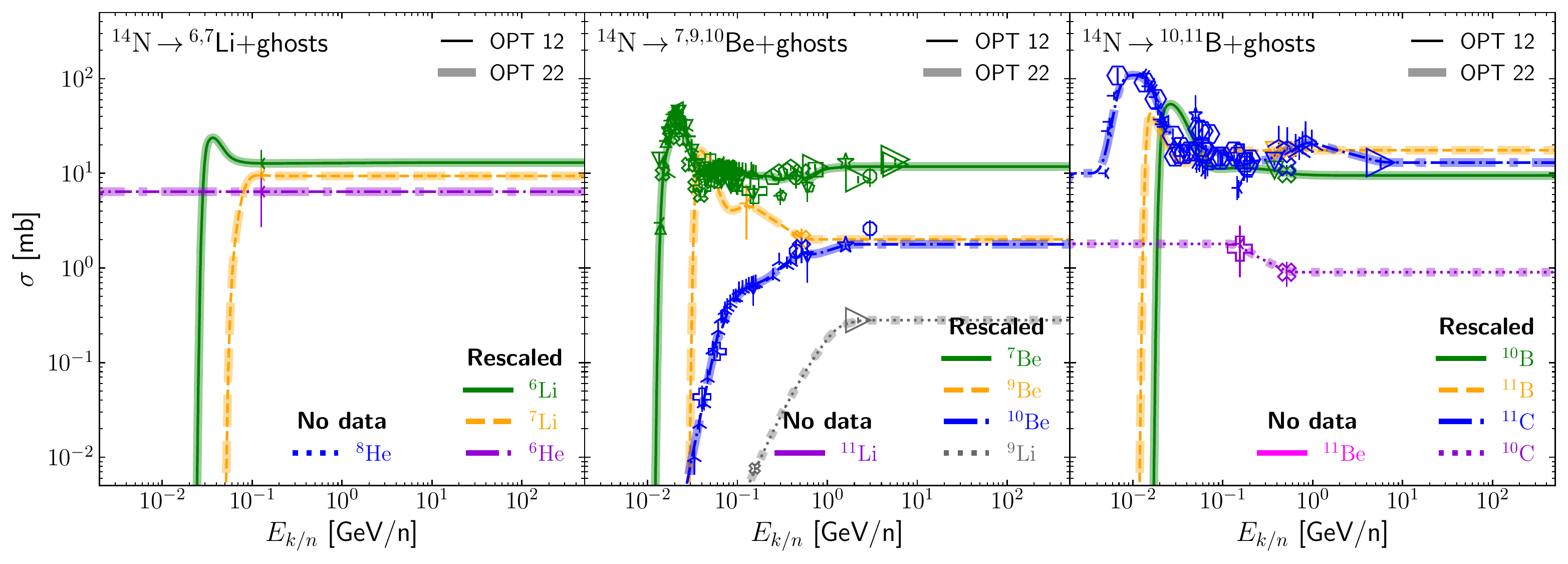}
   \includegraphics[trim={0 43 0 6},clip,width=0.81\textwidth]{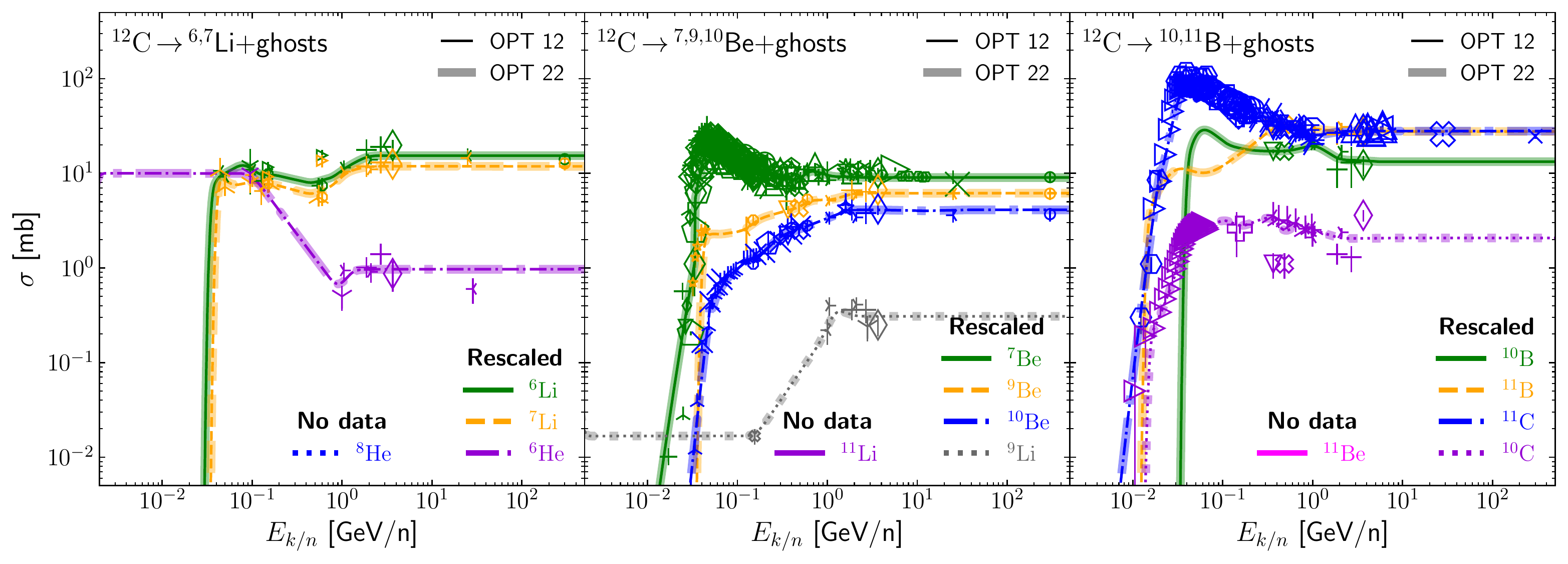}
   \includegraphics[trim={0 43 0 6},clip,width=0.81\textwidth]{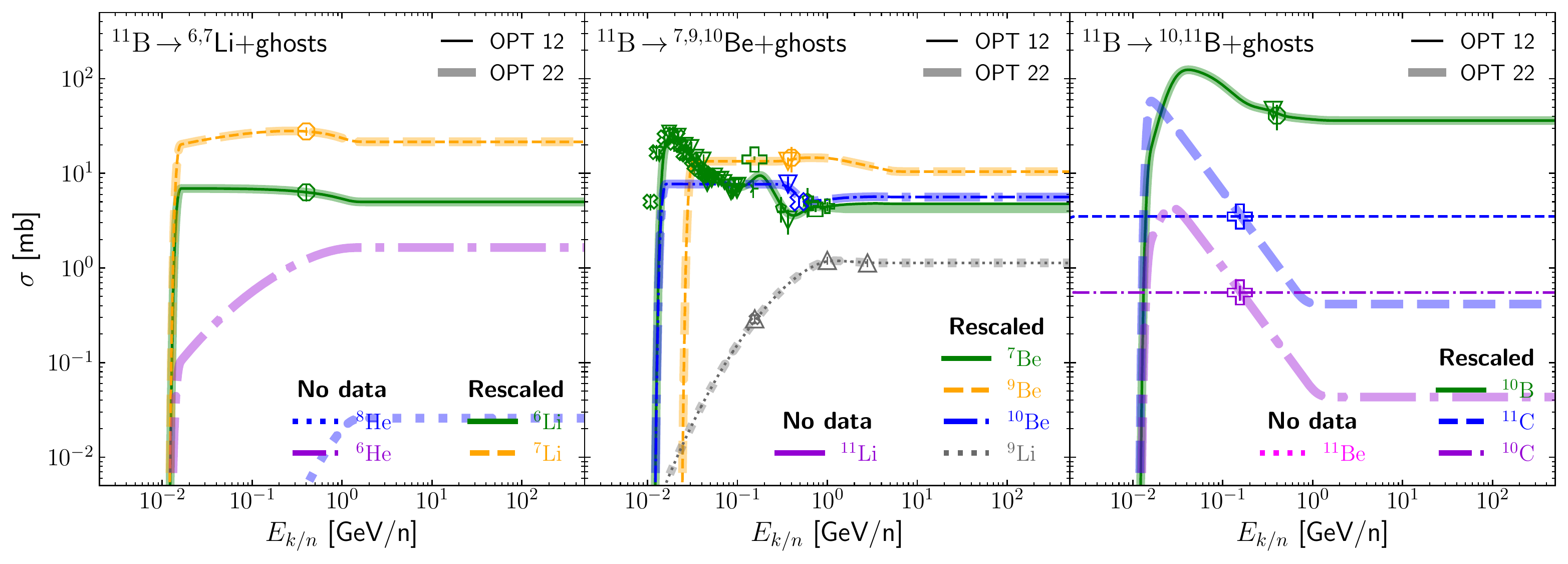}
   \includegraphics[trim={0 0  0 6},clip,width=0.81\textwidth]{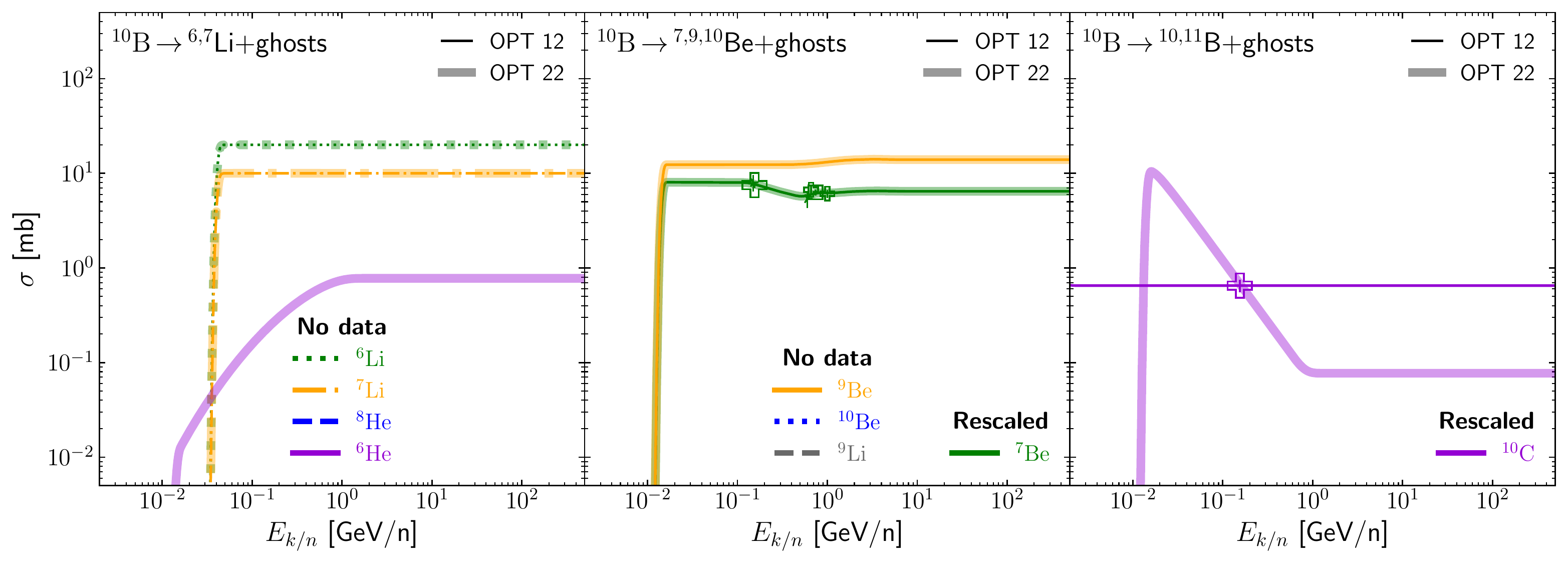}
   \caption{Same as \ref{fig:xs_data_vs_model_1}, but for  $^{15,14}$N, $^{12}$C, and $^{10,11}$B progenitors. The references for the data points are gathered in Fig.~\ref{fig:xs_data_vs_model_leg}.
   \label{fig:xs_data_vs_model_2}}
\end{figure*}
\begin{figure*}[h!] 
   \hspace{2.1cm}\includegraphics[trim={0 43 0 6},clip,width=0.81\textwidth]{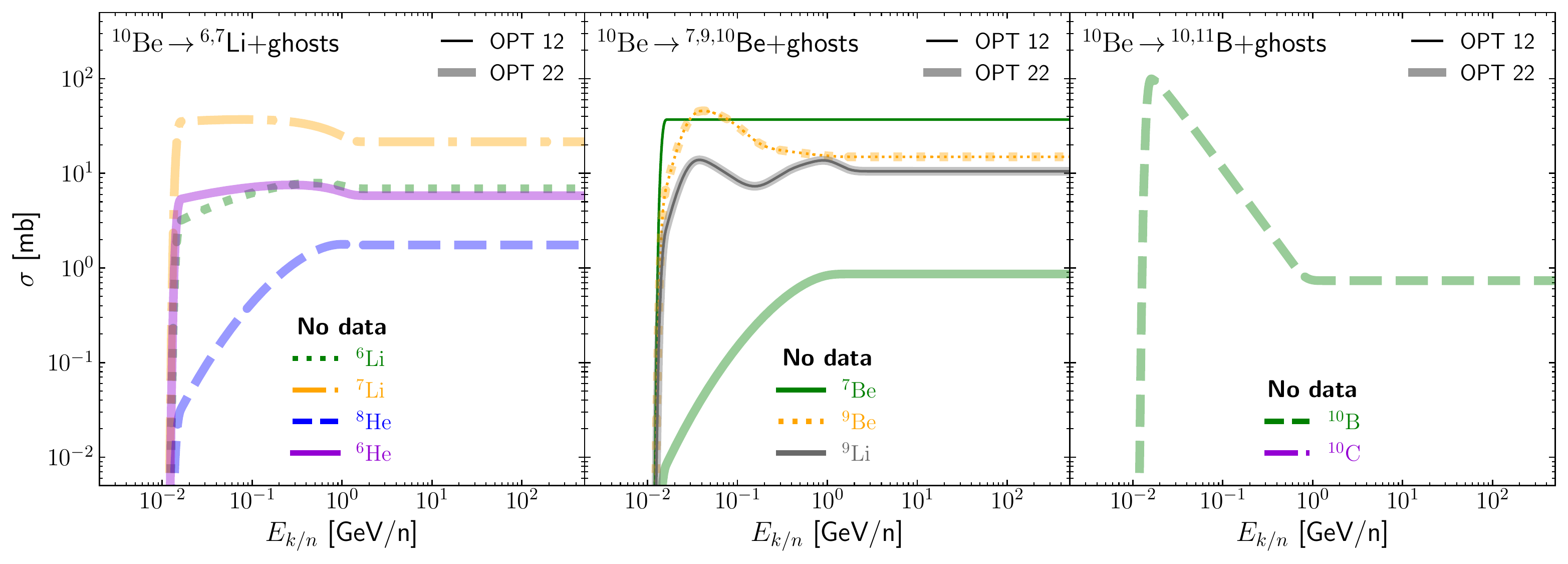}

   \hspace{2.1cm}\includegraphics[trim={0 43 6 6},clip,width=0.555\textwidth]{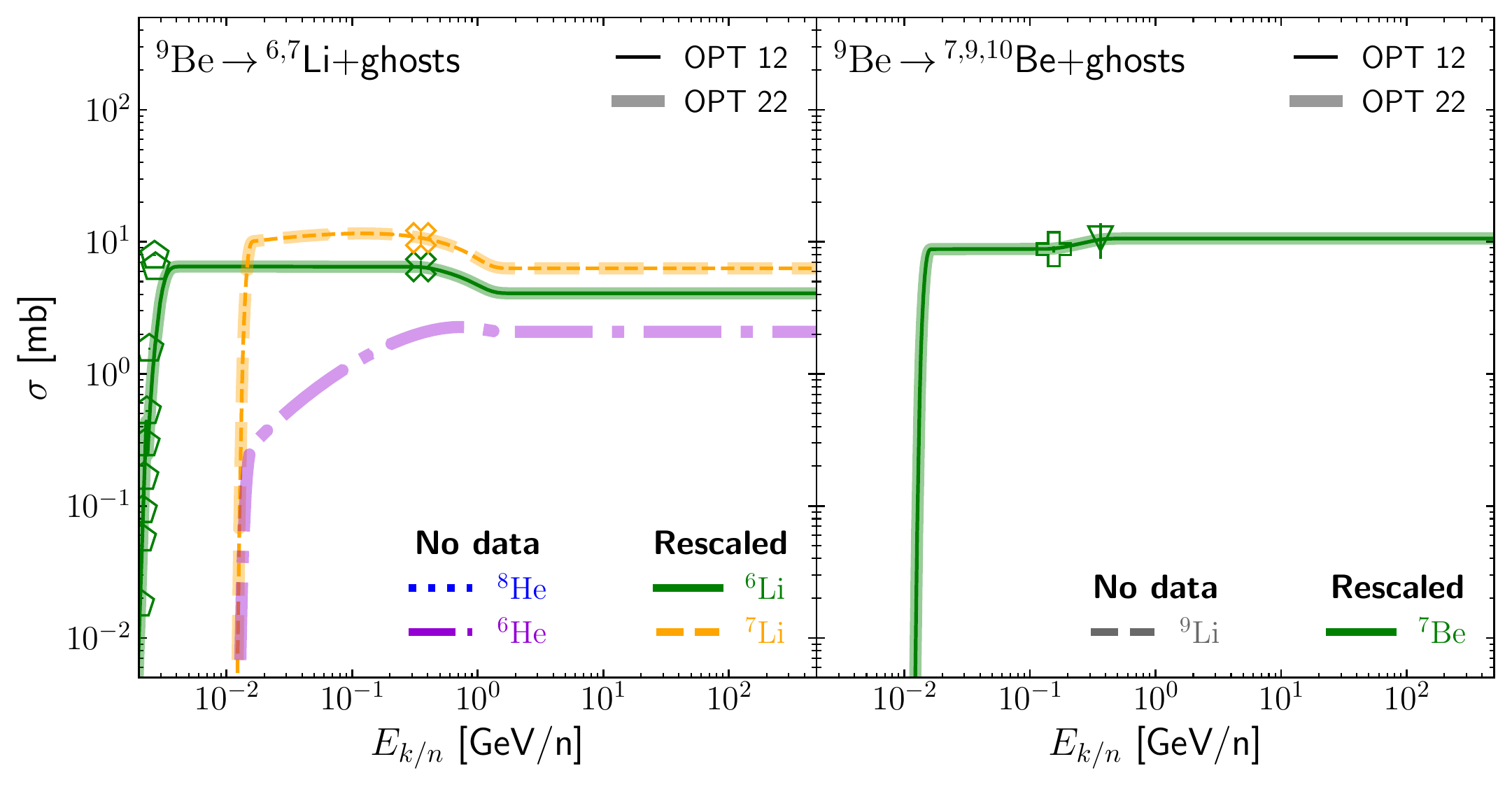}

   \hspace{2.1cm}\includegraphics[trim={0 43 0 6},clip,width=0.311\textwidth]{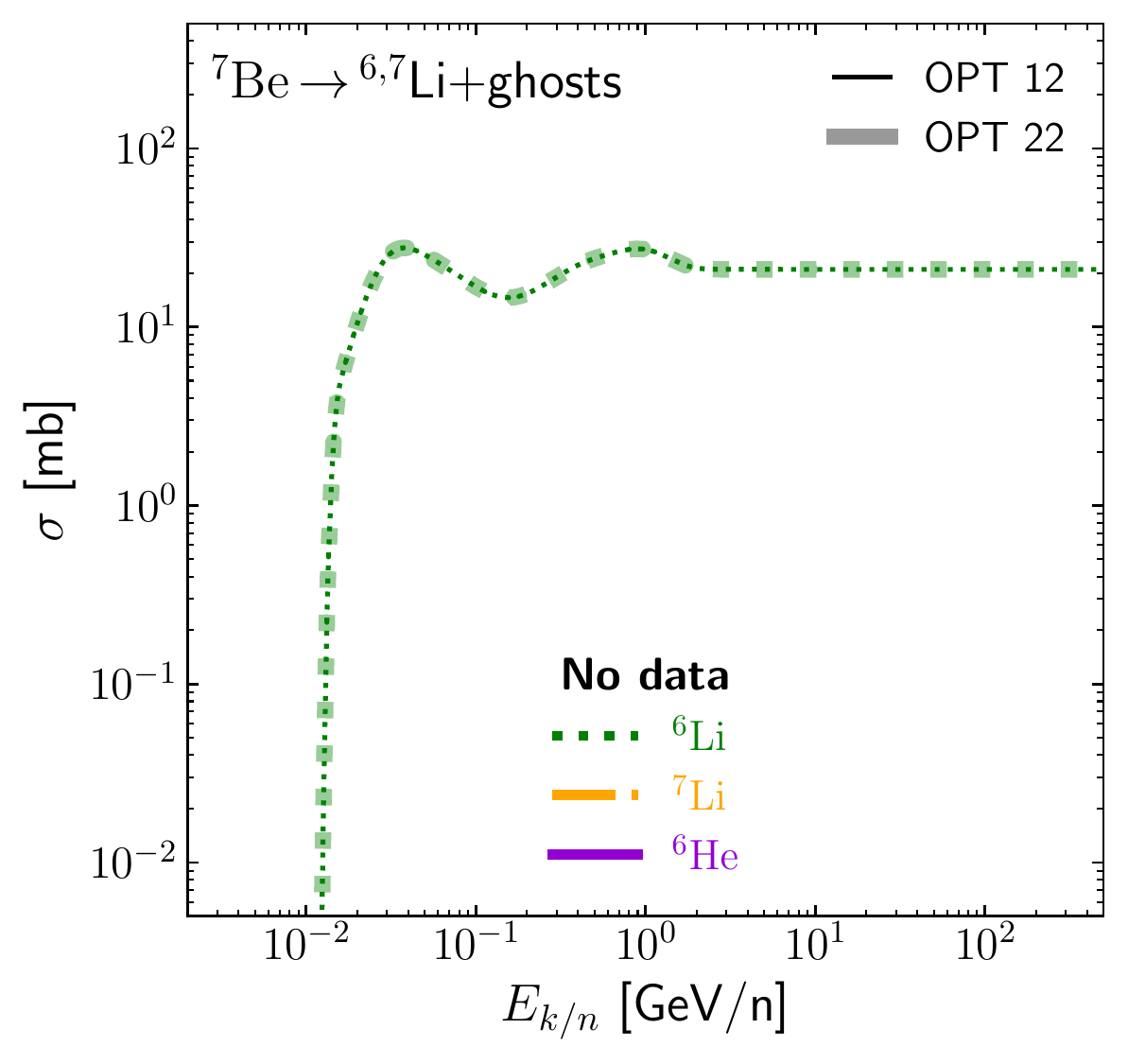}

   \hspace{2.1cm}\includegraphics[trim={0 0  0 6},clip,width=0.564\textwidth]{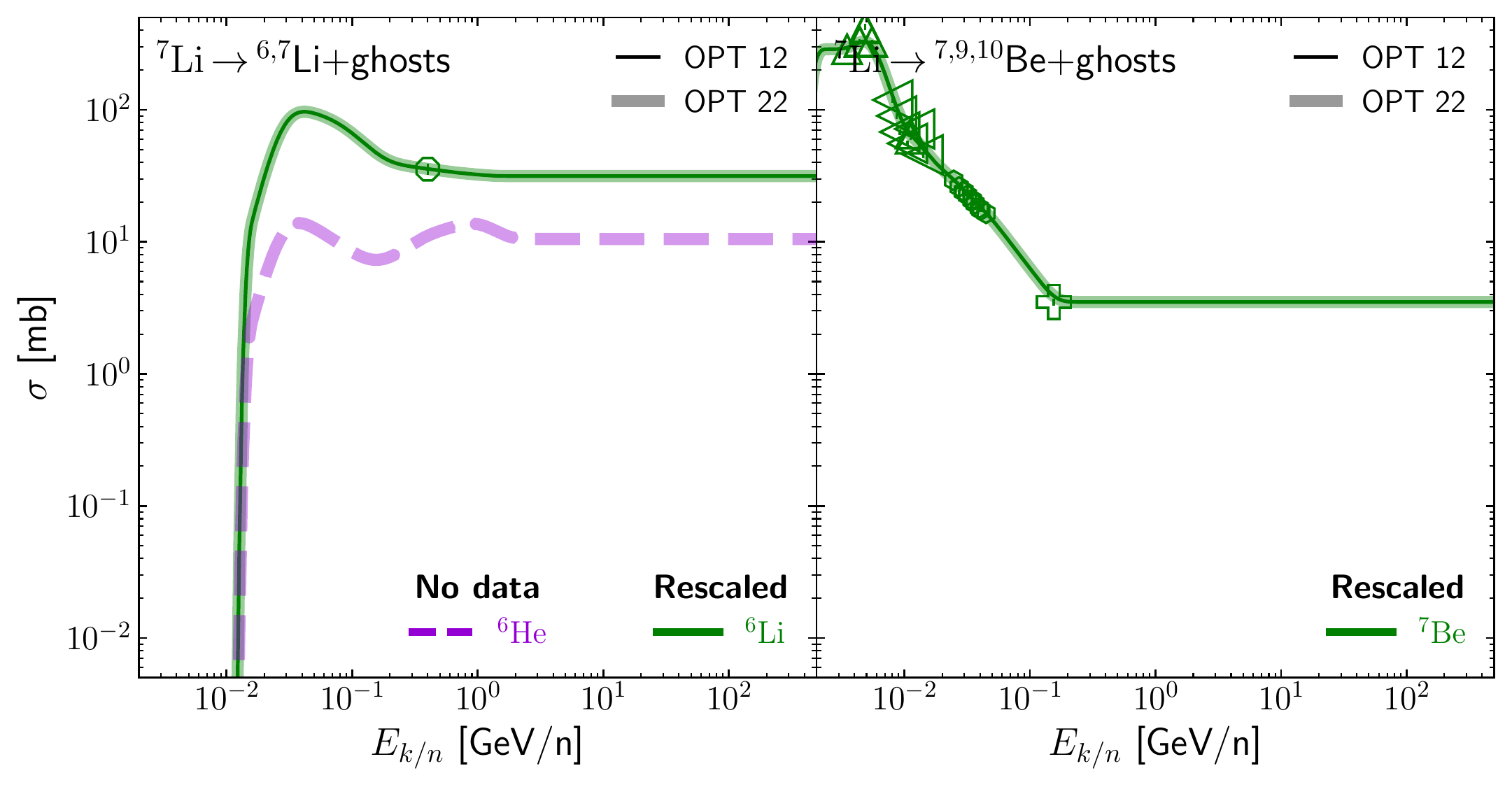}
   \caption{Same as \ref{fig:xs_data_vs_model_1}, but for and $^{10,9,7}$Be and  $^{7}$Li progenitors. The references for the data points are gathered in Fig.~\ref{fig:xs_data_vs_model_leg}.
   \label{fig:xs_data_vs_model_3}}
\end{figure*}
\begin{figure*}[t]
   \centering
   \includegraphics[width=1.\textwidth]{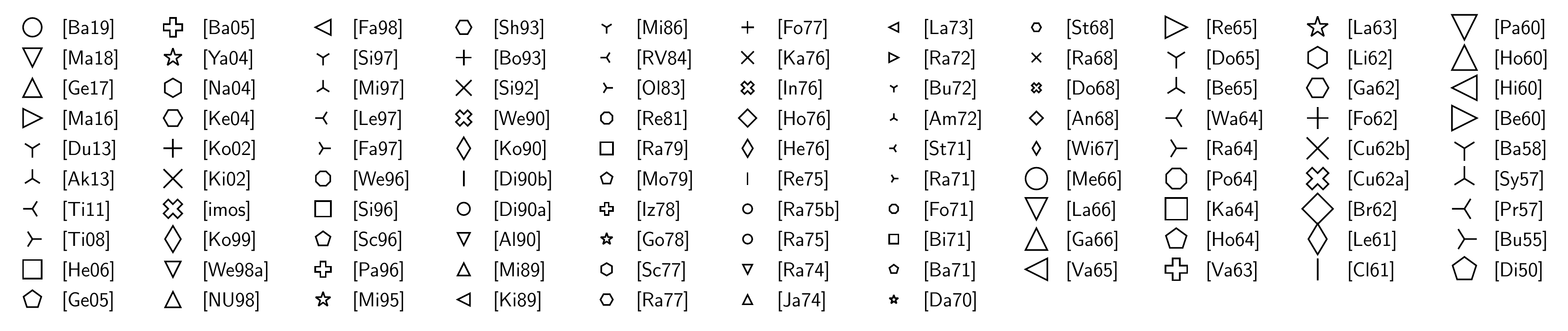}
   \caption{References for all data points shown in Figs.~\ref{fig:xs_data_vs_model_1}, \ref{fig:xs_data_vs_model_2}, and \ref{fig:xs_data_vs_model_3}, sorted from the most recent to the older ones. Specific labels from the original \galprop{} files are {\tt[imos]}, {\tt[We96]}, and {\tt[NU98]}, which correspond to Igor Moskalenko's additional data, Bill Webber private communication on \citet{1997ApJ...488..730R}, and data retrieved from Nuclex \citep{ivanov1998computer} respectively.
   \label{fig:xs_data_vs_model_leg}}
\end{figure*}
\nocite{2019NIMPB.454...50B} \nocite{2018NatSR...8.2570M} \nocite{Ge17} \nocite{2016NuPhA.946..104M} \nocite{2013PhRvC..88b4606D} \nocite{2013RadM...59..262A} \nocite{2011PAN....74..523T}
\nocite{2008PhRvC..78c4615T} \nocite{Ge05} \nocite{2006NIMPA.562..729H} \nocite{2005JETPL..81..140B} \nocite{2004PhRvC..70e4607N} \nocite{KETTERN2004939} \nocite{2004NIMPB.226..243Y} \nocite{2002NIMPB.196..239K}
\nocite{1999ICRC....4..267K} \nocite{1998ApJ...508..949W,1998PhRvC..58.3539W} \nocite{Fa98ref} \nocite{1997NIMPB.123..324S} \nocite{1997NIMPB.129..153M} \nocite{1997M&PSA..32Q..78L,1996NIMPB.113..429G} \nocite{fassbender1997activation} \nocite{1997NIMPB.123..324S} \nocite{1996NIMPB.114...91S} \nocite{1996NIMPB.113..470P} \nocite{1995NIMPB.103..183M} \nocite{1993PhRvC..48.2617S} \nocite{1993NIMPB..82....9B} \nocite{1992LPI....23.1305S} \nocite{1990PhRvC..41..566W} \nocite{1990NIMPA.291..662K} \nocite{1990NIMPB..52..588D,NSR1990DI06} \nocite{aleksandrov1990production}
\nocite{1989Ana...114..287M} \nocite{1989PhRvC..40...35K} \nocite{1986NIMPB..16...61M} \nocite{1984ADNDT..31..359R} \nocite{1983PhRvC..28.1602O} \nocite{1981E&PSL..53..203R}
\nocite{1979PhRvC..20..787R} \nocite{1979PhRvC..19..631M} \nocite{Iz78} \nocite{golikov1978cross} \nocite{1977NucIM.147..399S} \nocite{1977ICRC....2..203R} \nocite{1977PhRvC..15.2159F} \nocite{1976PhRvC..13..253K} \nocite{INOUE19761425} \nocite{1976PhRvC..14..753H} \nocite{1976PhRvC..14.1506H} \nocite{1975CRASB.280..513R} \nocite{1975PhRvC..12..915R} \nocite{1975PhLB...57..186R} \nocite{1974PhRvC...9.1385R} \nocite{1974PhRvC...9.2134J} \nocite{1973PhRvC...8..483L} \nocite{1972PhRvC...6..685R} \nocite{1972PhRvC...6.1153B} \nocite{1972NuPhA.195..311A} \nocite{1971NuPhA.175..124S} \nocite{1971PhRvL..27..875R} \nocite{1971NuPhA.165..405F} \nocite{bimbot1971spallation} \nocite{BARBIER19712720} \nocite{1970PhRvC...1..270D}
\nocite{1968NuPhB...5..188S} \nocite{RAYUDU19682311} \nocite{1968PhRv..169..836D} \nocite{ANDREWS1968689} \nocite{PhysRev.154.1005} \nocite{1966NucPh..78..476M} \nocite{doi:10.1139/v66-410} \nocite{NSR1966GA15} \nocite{1965NucPh..62...81V} \nocite{REEDER19651879} \nocite{1965PhRv..139.1513D} \nocite{1965PhL....15..147B} \nocite{warshaw1954cross} \nocite{doi:10.1139/v64-178} \nocite{1964PhRv..133.1507P} \nocite{doi:10.1139/p64-130} \nocite{1964NucPh..51..363H} \nocite{1963PhL.....7..163V} \nocite{1963JETP...16....1L} \nocite{1962PhRv..127.1269L} \nocite{1962NucPh..39..447G} \nocite{1962NucPh..31...43F} \nocite{1962PhRv..127..950C} \nocite{brun1962determination} \nocite{1961NucPh..25..216L} \nocite{1961PPS....78..681C} \nocite{1960NucPh..18..646P} \nocite{1960PhRv..118.1618H} \nocite{doi:10.1143/JPSJ.15.741} \nocite{1960PhRv..119..316B}
\nocite{PhysRev.112.1319} \nocite{Symonds_1957} \nocite{prokoshkin1957investigation} \nocite{Burcham_1955} \nocite{Dickson_1951}

\paragraph{Step 4: scaling for He targets.} Finally, not only cross sections on H matter, but cross sections on He ($\sim10\%$ in number in the ISM) are also necessary. As a last step, we rescale the re-evaluated cross sections on H using the empirical formulae of \citet{1988PhRvC..37.1490F}\footnote{See also \citet{2018JCAP...01..055R} for a discussion of possible shortcomings of this scaling relation.}. These formulae, however, if naively applied, lead to arbitrarily large enhancements for the extrapolation to small fragments (in particular Li) and low energies, so that we enforce an ad hoc maximal enhancement $(\sigma^{\rm He}/\sigma^{\rm H})_{\rm max} = 6$. This can be compared to the few data points of \citet{2004NIMPB.226..243Y}, for the fragmentation of $Z=24$ and $Z=26$ into $^7$Be: at 100 (resp. 230~MeV/n), $\sigma^{\rm He}/\sigma^{\rm H}=1.8$ (resp. $2.0$) for Cr, whereas $\sigma^{\rm He}/\sigma^{\rm H}=17$ for Fe\footnote{At 100 MeV/n, in order to calculate this ratio for Fe, we used for $\sigma^{\rm H}$ the data point from \citet{1997NIMPB.129..153M} because no corresponding data point was available from \citet{2004NIMPB.226..243Y}.} (resp. $11$). These measurements show irregularities that could probably not be captured by the scaling proposed by \citet{1988PhRvC..37.1490F}. However, as will be seen later, the contribution of Fe fragmentation for the production of Li and Be reaches at most $10\%$: even a factor 3 error on the production of He (which is only $10\%$ in number) would amount to a mere $\lesssim 3\%$ bias on the total production; moreover, these enhanced production of light fragments only applies below a few hundreds of MeV/n, that is much below the range where AMS-02 collects data. Nonetheless, they might be of interest to interpret very low-energy Voyager data \citep{2016ApJ...831...18C}.

\subsubsection{Updated cross sections and most impacted reactions}

As already said, we consider the most important CR progenitors for LiBeB production. We show the data and our new cross sections for the following CR isotopes: $^{56}$Fe, $^{28}$Si, $^{24}$Mg, $^{20}$Ne, and $^{16}$O progenitors in Fig.~\ref{fig:xs_data_vs_model_1}; then $^{14,15}$N, $^{12}$C, and $^{10,11}$B progenitors in Fig.~\ref{fig:xs_data_vs_model_2}; and finally $^{7,9,10}$Be and $^{7}$Li progenitors in Figs.~\ref{fig:xs_data_vs_model_3}. The legend and associated references for all the data presented are listed in Fig.~\ref{fig:xs_data_vs_model_leg}.

In Figs.~\ref{fig:xs_data_vs_model_1}, \ref{fig:xs_data_vs_model_2}, and \ref{fig:xs_data_vs_model_3}, the three columns (left, centre, and right) show the reactions related to the production of Li, Be, and B respectively. They include the direct production of stable (or CR-relevant) isotopes, but also the contributions from short-lived nuclei (dubbed ghosts), whose final decay state is one of these CR isotopes \citep{1984ApJS...56..369L,00/04/78/51/PDF/tel-00008773.pdf}: we recall that the cumulative cross-sections required for CR propagation corresponds to
\begin{eqnarray}
  \sigma^{\rm c}(X+H\to Y) &=& \sigma(X+H\to Y) 
  \label{eq:ghost} \\
  &+& \!\!\!\!\!\sum_{G\,\in\,{\rm ghosts}}\!\!\!\sigma(X+H\to G)\cdot {\cal B}r\,(G\!\to\!Y)\,,\nonumber
\end{eqnarray}
where ${\cal B}r\,(G\!\to\!Y)$ is the branching ratio for the ghost $G$ to end up as $Y$ (see the third row of Table~\ref{tab:LiBeB_XS_rescalingfactor_HE} for their values).
In each panel of Figs.~\ref{fig:xs_data_vs_model_1}--\ref{fig:xs_data_vs_model_3}, the data (symbols) and updated models (\optxii{} in thin lines and \optxxii{} in thick lines) are colour-coded according to the fragments considered. It is worth highlighting a few significant features: (i) when plenty of data exist, our procedure is very successful in matching the data at all energies; (ii) where only very few data exist, our procedure roughly amounts to a global rescaling of the initial cross sections; (iii) \xsGalxii{} and \xsGalxxii{} parametrisations can have very different energy dependences, and these differences remain in \optxii{} and \optxxii{} when only a few data are available (e.g. compare the thin and thick lines in the right-centre panel of Fig.~\ref{fig:xs_data_vs_model_1}, i.e. $^{24}$Mg in B through $^{11}$C); (iv) in several very relevant cases for which \xsGalxii{} was null, the updated cross section \optxii{} was replaced by a constant value passing through the few data points (e.g. top-left panel of Fig.~\ref{fig:xs_data_vs_model_1}, that is $^{56}$Fe in Li); (v) as already said, many reactions still lack any data (e.g. bottom-left panel of Fig.~\ref{fig:xs_data_vs_model_2}, that is $^{10}$Be in Li).
We finally stress that for many reactions, \xsGalxii{} and \xsGalxxii{} initial parametrisations were observed to be one and the same, so that their updated counterparts \optxii{} and \optxxii{} are also one and the same; this peculiarity comes from the fact that these initial parametrisations were probably already rescaled to pre-2004 nuclear data for their use in \galprop{}.

\begin{table*}[t]
\small
\centering
\caption{High-energy value of the rescaling factors ${\cal R}_{XS}^{method}=\sigma^{updated}_{XS}/\sigma^{init}_{XS}$ for the production of Li, Be, and B using a production cross-section set $XS$ built from $method$; we take $method$ here to be the rescaling with constant cross section above 1.5~GeV/n (dubbed `fit~$>\!1.5$~GeV/n', see Sect.~\ref{sec:procedure} and text for details). The first column shows the list of progenitors (or projectiles) and the first three lines show (i) the considered CR element, (ii) the associated list of fragments (isotopes and ghosts ${\cal B}{\rm r}>5\%$), and (iii) the branching ratios of the ghosts into CRs (in parenthesis). Entries in the table sometimes show two factors for the two updated $XS$ sets considered (${\cal R}_{XS=\,\rm \optxii{}}\,|\,{\cal R}_{XS=\,\rm \optxxii{}}$); however, most of the time a single number is shown, the original \xsGalxii{} and \xsGalxxii{} sets being similar for most reactions. The following special keys are also used: $\times$ means no-data available (hence no possible re-evaluation of the cross section), and $\infty$ refers to an ill-defined ${\cal R}$ (null value for $\sigma^{init}_{XS}$), though the updated cross-section value $\sigma^{updated}_{XS}$ is well defined.}
\label{tab:LiBeB_XS_rescalingfactor_HE}
\begin{tabular}{lccccccccccccccccc}
\toprule
                 & \multicolumn{4}{c}{Li} && \multicolumn{5}{c}{Be} && \multicolumn{6}{c}{B} \\
\cline{2-5}\cline{7-11}\cline{13-18}\\[-2.5mm]
                 & $^{6}{\rm Li}$ & $^{7}{\rm Li}$ & $^{6}{\rm He}$ &\!\!$^{8}{\rm He}\!\!$ && $^{7}{\rm Be}$ & $^9{\rm Be}$ & $^{10}{\rm Be}$ &\!\!$^{11}{\rm Li}$\!\!& $^9{\rm Li}$ && $^{10}{\rm B}$ & $^{11}{\rm B}$ & $^{11}{\rm C}$ & $^{10}{\rm C}$ & $^{11}{\rm Be}$&\!\!$^{11}{\rm Li}$\!\!\\
                 &                &                &\!\!\!\!(100\%)\!\!\!\!&\!\!(16\%)\!\!      &&                &                &                 &\!\!\!\!(85\%)\!\!\!\!   &\!\!\!(49\%)\!\!\! &&           &                &\!\!\!\!(100\%)\!\!\!\!&\!\!\!\!(100\%)\!\!\!\!&\!\!\!(97\%)\!\!\!&\!\!\!\!(7.8\%)\!\!\!\!\\
\midrule
 $^{56}{\rm Fe}$\!&   $\infty$|0.8 &     $\infty$|1 &   $\infty$|0.6 &       $\times$ &&         15|0.8 &         21|1.4 &          19|0.7 &        $\times$ &   $\infty$|0.3 &&         20|0.8 &           15|1 &        2.0|1.7 &       $\times$ &        $\times$ &        $\times$ \\
 $^{28}{\rm Si}$\!&       $\times$ &       $\times$ &       $\times$ &       $\times$ &&              1 &           1.05 &            1.02 &        $\times$ &            0.4 &&       $\times$ &       $\times$ &        0.5|1.2 &       $\times$ &        $\times$ &        $\times$ \\
 $^{24}{\rm Mg}$\!&       $\times$ &       $\times$ &       $\times$ &       $\times$ &&           1.04 &           2.04 &            0.95 &        $\times$ &            0.6 &&       $\times$ &       $\times$ &        0.5|1.1 &       $\times$ &        $\times$ &        $\times$ \\
 $^{20}{\rm Ne}$\!&       $\times$ &       $\times$ &       $\times$ &       $\times$ &&       $\times$ &       $\times$ &        $\times$ &        $\times$ &       $\times$ &&       $\times$ &       $\times$ &           0.95 &              1 &        $\times$ &        $\times$ \\
  $^{16}{\rm O}$\!&            1.4 &           0.96 &       $\infty$ &       $\times$ &&           0.98 &           1.41 &            1.18 &        $\times$ &            0.7 &&           0.96 &            0.4 &       $\infty$ &       $\infty$ &        $\infty$ &        $\times$ \\
  $^{15}{\rm N}$\!&              1 &              1 &       $\times$ &       $\times$ &&              1 &              1 &               1 &        $\times$ &            0.5 &&           1.34 &           1.17 &              1 &       $\times$ &        $\times$ &        $\times$ \\
  $^{14}{\rm N}$\!&              1 &              1 &       $\infty$ &       $\times$ &&           1.18 &           0.94 &            1.02 &        $\times$ &            0.8 &&           0.91 &            0.6 &       $\infty$ &       $\infty$ &        $\times$ &        $\times$ \\
  $^{12}{\rm C}$\!&            1.1 &           0.94 &       $\infty$ &       $\times$ &&           0.94 &           0.91 &            1.04 &        $\times$ &       $\infty$ &&           1.08 &           0.92 &           1.04 &            0.7 &        $\times$ &        $\times$ \\
  $^{11}{\rm B}$\!&              1 &              1 &       $\times$ &       $\times$ &&\!\!1.06|1.16\!\!&          1.04 &             0.4 &        $\times$ &           0.97 &&           0.93 &          \dots &   $\infty$|0.7 &  $\infty$|1.16 &        $\times$ &        $\times$ \\
  $^{10}{\rm B}$\!&       $\times$ &       $\times$ &       $\times$ &       $\times$ &&           0.94 &       $\times$ &        $\times$ &           \dots &       $\times$ &&          \dots &          \dots &          \dots &  $\infty$|1.75 &           \dots &           \dots \\
 $^{10}{\rm Be}$\!&       $\times$ &       $\times$ &       $\times$ &       $\times$ &&       $\times$ &       $\times$ &           \dots &           \dots &       $\times$ &&       $\times$ &          \dots &          \dots &       $\times$ &           \dots &           \dots \\
  $^9{\rm Be}$\!&              1 &              1 &       $\times$ &       $\times$ &&              1 &          \dots &           \dots &           \dots &       $\times$ &&          \dots &          \dots &          \dots &          \dots &           \dots &           \dots \\
  $^{7}{\rm Be}$\!&       $\times$ &       $\times$ &       $\times$ &          \dots &&          \dots &          \dots &           \dots &           \dots &          \dots &&          \dots &          \dots &          \dots &          \dots &           \dots &           \dots \\
  $^{7}{\rm Li}$\!&              1 &          \dots &       $\times$ &          \dots &&       $\infty$ &          \dots &           \dots &           \dots &          \dots &&          \dots &          \dots &          \dots &          \dots &           \dots &           \dots \\
\bottomrule
\end{tabular}
\end{table*}
In order to have a more concise view of the main changes in these derived cross sections, we gather in Table~\ref{tab:LiBeB_XS_rescalingfactor_HE} the ratio of the updated to initial cross sections above 1.5~GeV/n, ${\cal R}=\sigma^{\rm updated}/\sigma^{\rm initial}$, that is in a regime where we enforce the cross sections to be constant. The first column and first row respectively show the progenitors and fragments that appear in Figs.~\ref{fig:xs_data_vs_model_1}--\ref{fig:xs_data_vs_model_3}. The main comments we can make on this table are: (i) while data exist for the most important LiBeB contributors (C, N, and O, see Sect.~\ref{sec:xs_ranking}), the less important Si, Mg, and Ne channels have no data at all for the production of Li and $^{10,11}$B; (ii) contributions from the secondary species $^{10}$B and $^7$Be into Li---that peak at GeV/n energies, see \citet{2018PhRvC..98c4611G}---also lack data; (iii) there is also no data for the production of $^8$He and $^{11}$Li ghost nuclei, however, owing to their small branching ratios and small cross-section predictions (compared to the other isotopes), these reactions are not expected to even reach the percent level of contribution overall; (iv) surprisingly, the initial cross sections \xsGalxii{} and \xsGalxxii{} for the fragmentation of $^{12}$C and $^{16}$O into the short-lived $^{10}$C and $^{11}$C nuclei were null (a closer inspection indicates that these contributions were probably added to the direct production of $^{10}$Be and $^{11}$B; (v) the most striking numbers are those for the fragmentation of $^{56}$Fe, indicating that the initial \xsGalxii{} (left-hand side numbers in the cells) significantly undershot the data, whereas the initial \xsGalxxii{} (right-hand side numbers in the cells) predictions were more or less in line with the data (with an overall small overshoot though). Looking back at Fig.~\ref{fig:impact_heaviest}, this latter behaviour seems to confirm and favour the case for an enhanced production of Li with regard to both Be and B (dotted lines).

\begin{figure}[t]
   \centering
   \includegraphics[trim={7 0 6 6},clip,width=0.54\columnwidth]{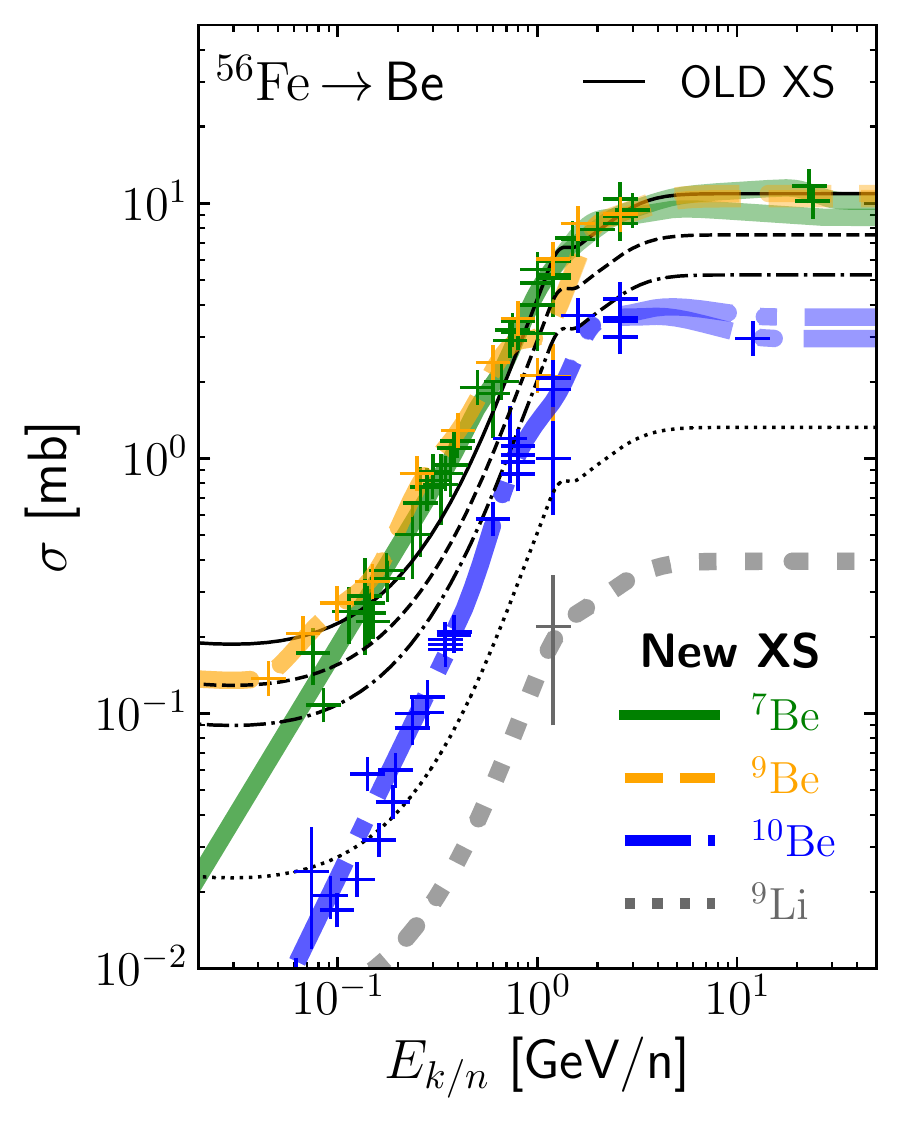}
   \includegraphics[trim={56 0 0 6},clip,width=0.446\columnwidth]{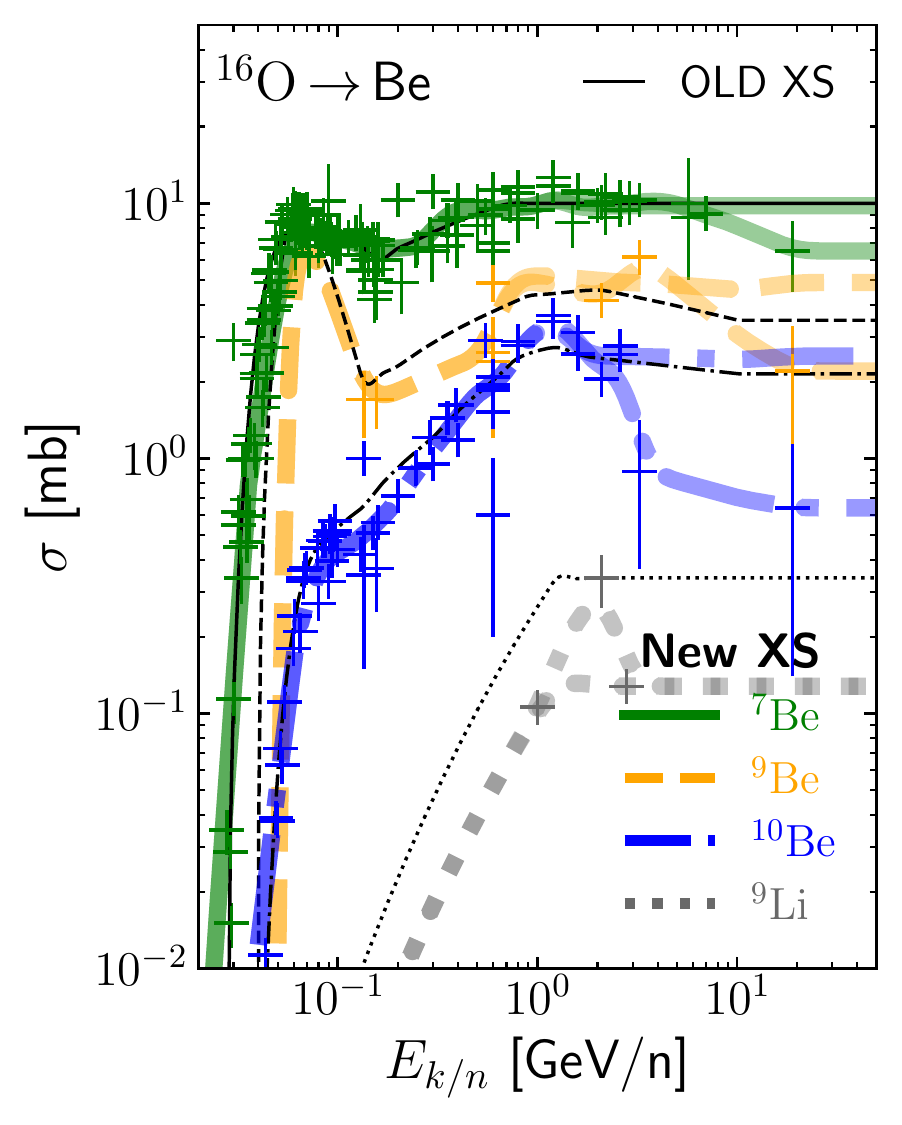}
   \caption{Comparison of original (\xsGalxxii{}, thin black lines labelled `OLD XS') and updated (\optxxii{}, thick coloured lines labelled `NEW XS') production cross sections, for $^{56}$Fe (left) and $^{16}$O (right) fragmenting into Be isotopes. The two branches seen in some of the updated cross sections above 1.5 ~GeV/n illustrate two different procedures to update them (fit of a constant or spline passing through the data, see Sect.~\ref{sec:procedure}). The ratio of the values in these two branches at high energy corresponds to the values $\tilde{\cal R}={\cal R}^{{\rm fit}~>1.5~{\rm GeV/n}}/{\cal R}^{\rm no-fit}$ shown in Table~\ref{tab:LiBeB_XS_rescalingfactor_HE_comp}.
   \label{fig:xs_old_vs_new}}
\end{figure}

\begin{table}[t]
\small
\centering
\caption{Ratio of high-energy rescaling factors between two methods tested, $\tilde{\cal R}={\cal R}^{{\rm fit}~>1.5~{\rm GeV/n}}/{\cal R}^{\rm no-fit}$, with `fit~$>\!1.5$~GeV/n' for which the cross-section is constant above 1.5~GeV/n (fit on existing data and their errors) and `no-fit' for which the model goes through all available points. First column (projectiles) and line (fragments) are as in Table~\ref{tab:LiBeB_XS_rescalingfactor_HE}. Otherwise, we only show here reactions for which the two re-scaling methods used in $\tilde{\cal R}$ differ by more than $5\%$.}
\label{tab:LiBeB_XS_rescalingfactor_HE_comp}
\begin{tabular}{lccccccc}
\toprule
                & \multicolumn{4}{c}{Be} && \multicolumn{2}{c}{B} \\
                \cline{2-5}\cline{7-8}\\[-2.5mm]
                 & $^{7}{\rm Be}$ & $^9{\rm Be}$ & $^{10}{\rm Be}$ & $^9{\rm Li}$ && $^{10}{\rm B}$& $^{10}{\rm C}$\\
                 &                &                &                 &    (49\%)      &&               &       (100\%)  \\
\midrule
 $^{56}{\rm Fe}$ &            0.9 &          0.9   &          1.2    &         \dots  &&        \dots  &         \dots  \\
  $^{16}{\rm O}$ &            1.4 &          2.4   &          3.9    &           1.75 &&         0.67  &         \dots  \\
  $^{14}{\rm N}$ &            0.8 &         \dots  &          0.7    &         \dots  &&        \dots  &         \dots  \\
  $^{12}{\rm C}$ &         \dots  &         \dots  &          1.1    &         \dots  &&         1.11  &           0.6  \\
\bottomrule
\end{tabular}
\end{table}
Table~\ref{tab:LiBeB_XS_rescalingfactor_HE_comp} shows the ratio $\tilde{\cal R}={\cal R}^{{\rm fit}~>1.5~{\rm GeV/n}}/{\cal R}^{\rm no-fit}$, that is the comparison of two different choices for the high-energy update of the cross-section parametrisations: the numerator, ${\cal R}^{{\rm fit}~>1.5~{\rm GeV/n}}$, corresponds to the rescaling factors shown in Table~\ref{tab:LiBeB_XS_rescalingfactor_HE} where cross sections are enforced to be constant above $1.5$~GeV/n; the denominator, ${\cal R}^{\rm no-fit}$, corresponds to the case in which the cross sections follow closely all structures seen in the high-energy data (see Sect.~\ref{sec:procedure}). This $\tilde{\cal R}$ factors gives a hint of the variability of the data at high energy, and how sensitive the cross-section modelling is to these data. For readability, Table~\ref{tab:LiBeB_XS_rescalingfactor_HE_comp} only shows values significantly different from 1, that is reactions directly sensitive to this rescaling choice. We see that the fragmentation of $^{56}$Fe into Be isotopes and ghosts, and also that of $^{16}$O into Be and $^{10}$B, are significantly dependent on the two renormalisation procedure used. For a clearer illustration of this dependence, we show in Fig.~\ref{fig:xs_old_vs_new} the production of Be from $^{56}$Fe (left panel) and $^{16}$O (right panel). The updated cross sections (thick coloured lines) clearly split above 1.5~GeV/n, and the case of $\sigma^{\mathrm{^{16}O+H\to^{10}Be}}$ illustrates the difficulty to favour one or the other approach, as the high-energy points are clearly below those at $\sim 1$~GeV/n, though with much larger error bars. We also show on these plots the initial cross sections (thin black lines), to highlight the fact that moving from the old to the new cross sections is often more than a mere rescaling.

\subsection{Impact on Li/C, Be/C, and B/C ratios}
\label{sec:xs_impact}

To start this subsection, we recall that we considered two different initial cross-section parametrisations, namely \xsGalxii{} and \xsGalxxii{}, rescaled to nuclear data and updated into \optxii{} and \optxxii{}. We also build a third cross-section set, where cross sections without data are taken from \xsGalxii{}, while those with data and updated are taken from \optxxii{}; this hybrid configuration is denoted \optxiiupxxii{}\footnote{The motivation behind this set is that \xsGalxxii{} provides a more realistic energy dependence for Fe fragmentation (\xsGalxii{} has null values), though it is not clear, for reactions without data, whether \xsGalxxii{} or \xsGalxii{} is more appropriate.}.

\begin{figure}[t]
  \includegraphics[width=\columnwidth]{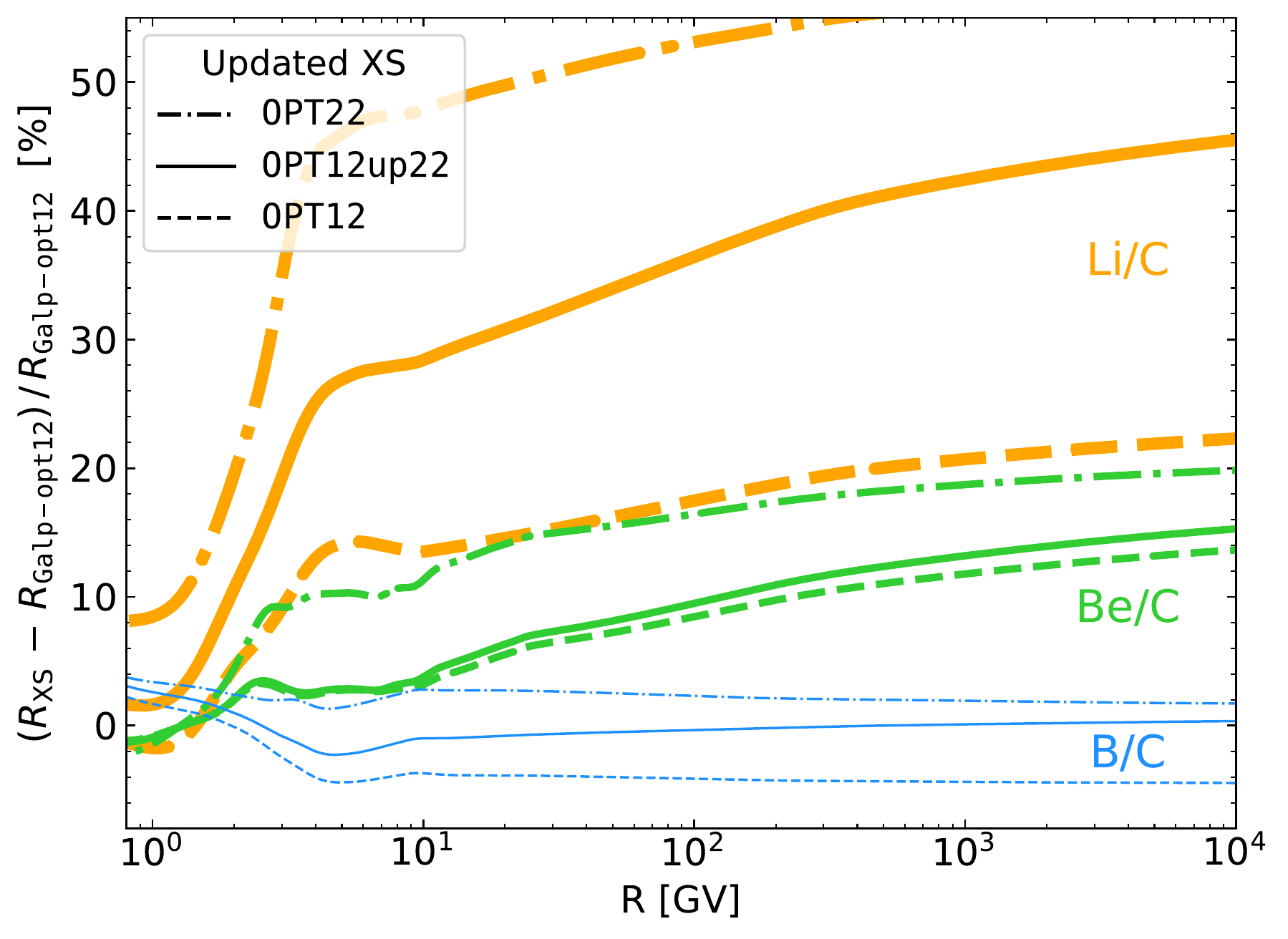}
  \caption{Enhancement of Li/C (orange thick lines), Be/C (green lines), and B/C (blue thin lines) flux ratios with regard to calculations performed using \xsGalxii{} (used in all recent studies). The three sets of curves correspond to the updated production cross-section sets \optxii{} (dashed lines), \optxiiupxxii{} (solid lines), and \optxxii{} (dash-dotted lines).
  \label{fig:impact_xsupdate}}
\end{figure}

We can now provide and bracket the impact these updated production cross sections have for the calculation of CR LiBeB/C ratios, as shown in Fig.~\ref{fig:impact_xsupdate}. These curves tell us that heavy species up to Fe matter for the LiBeB production (see Fig.~\ref{fig:impact_heaviest}). As can be seen, whereas B is not impacted ($\lesssim4\%$) by the use of the updated cross-section sets, both Be and Li are significantly increased with regard to calculations performed with the original set \xsGalxii{}. These enhancements are also more important for Li than for Be, and this ordering holds true for the three updated cross-section sets. To better understand the origin of the increased production of Li and Be, we show in App.~\ref{app:impact_per_progenitor} the impact of individual progenitors. It confirms that Fe updated cross sections are the main driver of the Li and Be enhancements; the update of O and Si cross-sections is also contributing to this enhancement. More precisely, we can identify three different origins for the changes in the predicted (LiBeB)/C ratios, with the combined impact of (e.g. at 10~GV): (i) accounting for $^{56}$Fe fragmentation when moving from \xsGalxii{} to \xsGalxxii{} that can be read off Fig.~\ref{fig:impact_heaviest} ($+15\%$ for Li, $+10\%$ for Be, and $+4\%$ for B); (ii) renormalising \xsGalxxii{} into \optxxii{} for $^{56}$Fe reactions and other progenitors as read off the right panel of Fig~\ref{fig:impact_xsrescaled_ratios} ($+6\%$ for Li, $+2\%$ for Be, and $-4\%$ for B); (iii) accounting for the differing cross sections values for unmeasured reactions in \xsGalxii{} and \xsGalxxii{}. The latter is mostly significant for Li, as read off Fig.~\ref{fig:impact_xsupdate}, indicating that the global change moving from \xsGalxii{} to \optxiiupxxii{} can be even larger than the sum of the two first effects ($+50\%$ for Li, $+10\%$ for Be, and $+2\%$ for B).

Concerning the spread between the \optxii{}, \optxiiupxxii{}, and \optxxii{} sets in Fig.~\ref{fig:impact_xsupdate}, we observe that the enhancement for Be is relatively well predicted ($6\%$ spread between the updated cross-section sets), while the spread on Li remains quite large ($\sim10-20\%$). This spread originates from the different values of the cross sections in the three updated sets. Indeed, for Fe, \optxxii{} (dash-dotted lines) provides a growing production cross sections with energy, whereas it is constant in \optxii{} (see top left panel of Fig.~\ref{fig:xs_data_vs_model_1}). The difference between \optxiiupxxii{} (solid lines) and \optxxii{} (dash-dotted lines) is more subtle and related to cross sections not rescaled: in the former set based on \xsGalxxii{} parametrisation, these cross sections are globally larger than in the latter set based on \xsGalxii{}. It is not clear whether the observed spread could be reduced by also applying our procedure to less-important cross-section reactions, or whether this difference is related to still unmeasured reactions; this certainly deserves further inspection.
Finally, it is also difficult to decide which of the three updated cross-section sets proposed is the most motivated, all the more because, for the few reactions studied here and that were not renormalised to the data before, the cross sections from \xsGalxii{} were found on average to undershoot the data, while those from \xsGalxxii{} to overshoot them (see Table~\ref{tab:LiBeB_XS_rescalingfactor_HE}).

\section{Updated ranking for Li production}
\label{sec:xs_ranking}

As underlined several times already, a thorough study and rankings of the most important reactions (or projectiles) involved in the production of Li to N fluxes were provided in \citet{2018PhRvC..98c4611G}. Owing to the important role of Fe fragmentation---that was not appreciated at that time---these ranking need to be updated. Instead of directly updating Table~V-VII and X-XII of \citet{2018PhRvC..98c4611G}, we provide a complementary view here, grouping contributions of specific CR elements. In words, for any given charge of a CR element $Z_i$, we can track all the contributions leading directly or indirectly to Li, Be, or B. Mathematically-speaking, we can define the $f_{i\to j}$ coefficients, which correspond to the fractional contribution of a CR isotope $i$ ending up in a CR isotope $j$, via direct (1-step) or multi-step production. These coefficients read (see Eq.~A1 in \citealt{2018PhRvC..98c4611G})
\begin{equation}
  f_{i\to j} =  f^{\rm 1-step}_{ij} + \sum_{i>k>j} f^{\rm 2-step}_{ikj} + \sum_{i>k>l>j} f^{\rm 3-step}_{iklj} + \dots,
\end{equation}
where $i>k$ means that the CR isotope $i$ is heavier than $k$.
Denoting $\sigma_{ij}$ the cross section of $i$ fragmenting in $j$ (for any ISM target), the $f_{i\to j}$ coefficients can be calculated from
\begin{equation}
  f_{i\to j} = \frac{\psi_j^{\rm sec}\big(\sigma_{kk'}\!=\!0\;{\rm for}\;k>i\,\&\,k'\leq k\big)-\psi_j^{\rm sec}\big(\sigma_{ik}\!=\!0\;{\rm for}\;k\leq i\big)}{\psi_j^{\rm sec}}\;,
\end{equation}
which by construction, ensures $\sum_i f_{i\to j} = 1$.
We can then define the fractional contribution $f_{Z_i\to Z_j}$ of a CR element of charge $Z_i$ into a CR element of charge $Z_j$ (with $n_{Z_j}$ isotopes) to be
\begin{equation}
  f_{Z_i\to Z_j} =  \frac{1}{n_{Z_j}}\times \displaystyle\sum_{\forall Z(i)=Z_i,\,\forall Z(j)=Z_j} f_{i\to j}\,,
\end{equation}
which also satisfies $\sum_{Z_i} f_{Z_i\to Z_j}=1$.

\begin{figure}[t]
  \includegraphics[trim={7 48 6 15},clip,width=\columnwidth]{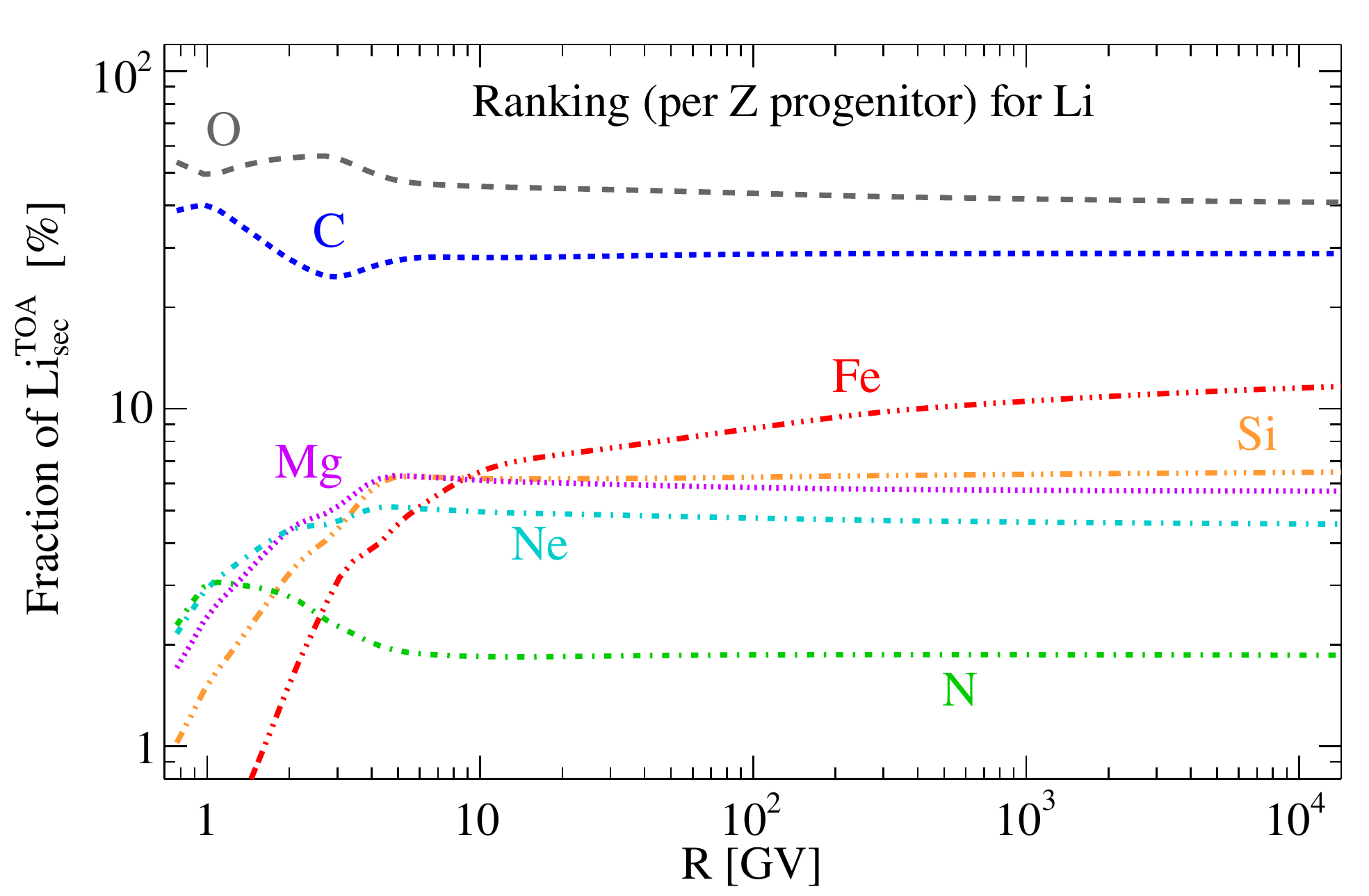}
  \includegraphics[trim={7 48 6 15},clip,width=\columnwidth]{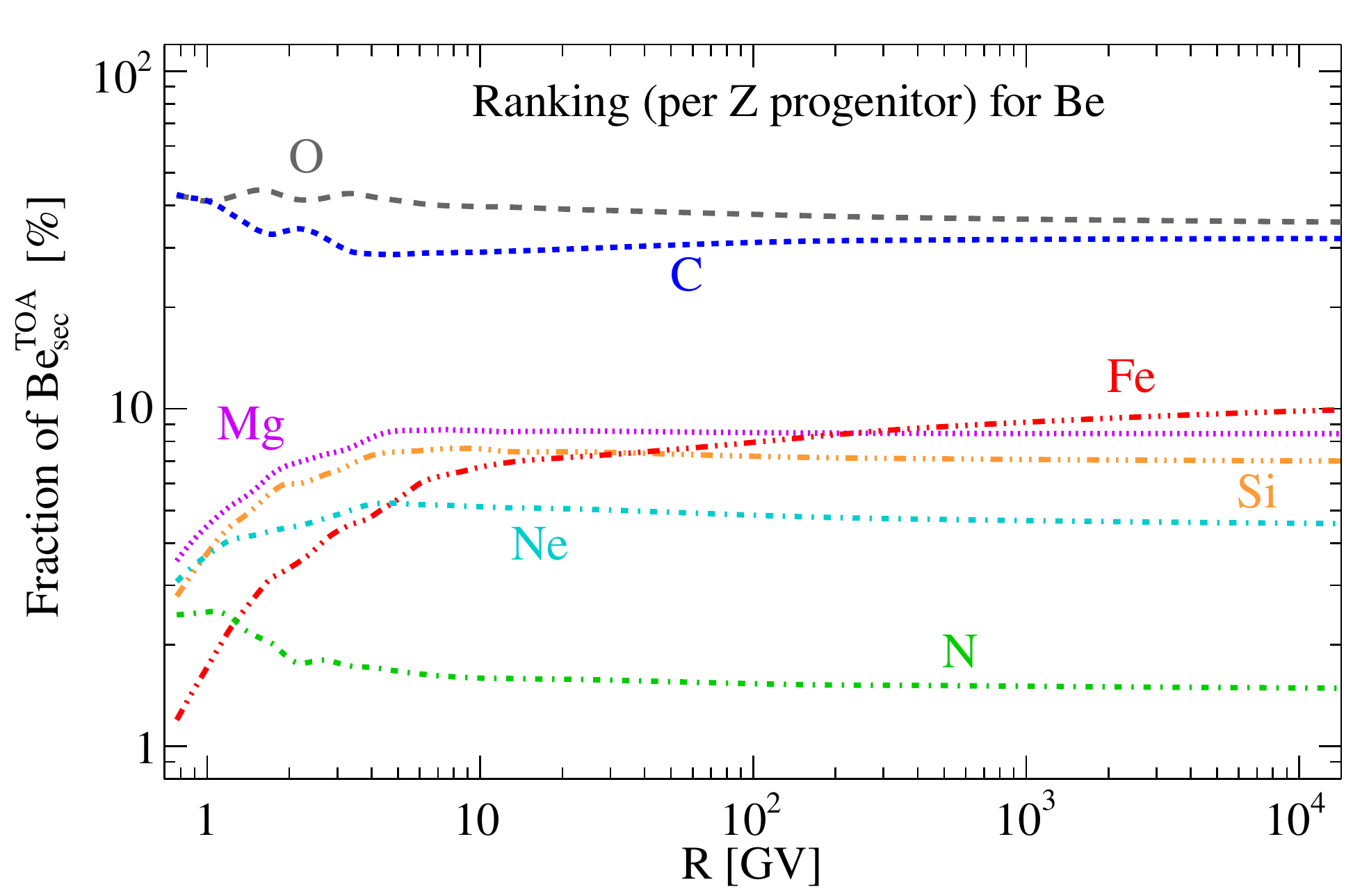}
  \includegraphics[trim={7 0 6 15},clip,width=\columnwidth]{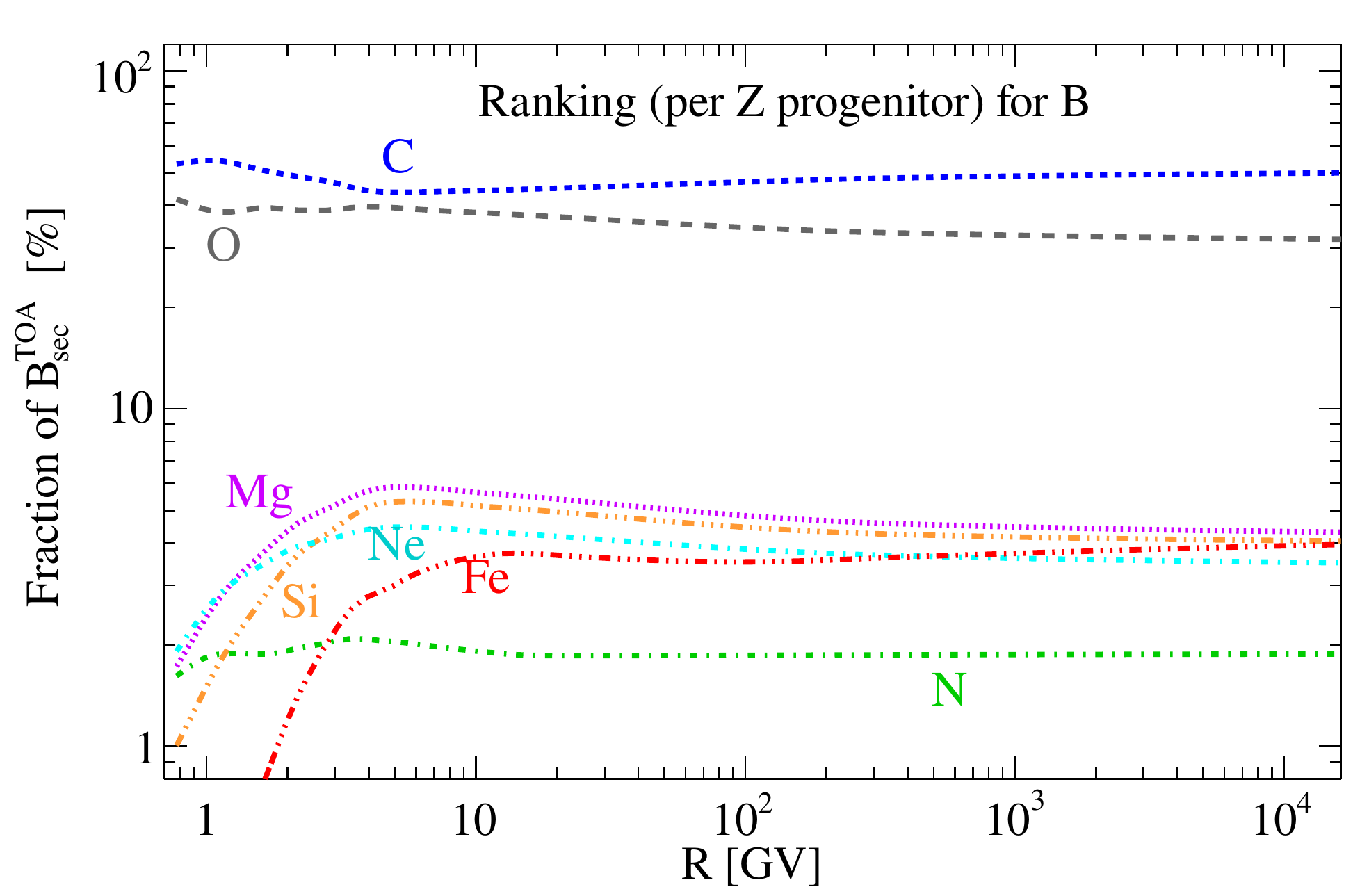}
  \caption{Colour-coded ranked progenitors (elements) of the Li, Be, and B (from top to bottom) fluxes as a function of rigidity. Calculations are based on \optxiiupxxii{} production cross-section set with fluxes modulated at 700~MV. For readability, only contributing fractions larger than $1\%$ are shown.
  \label{fig:ranking_per_element}}
\end{figure}
Figure~\ref{fig:ranking_per_element} shows from top to bottom, using \optxiiupxxii{}, the quantities $f_{Z_i\to {\rm Li}}$, $f_{Z_i\to {\rm Be}}$, and$f_{Z_i\to {\rm B}}$ as a function of rigidity. Only progenitor elements having contributing fractions $>1\%$ are shown, in order not to overcrowd the plot. As in \citet{2018PhRvC..98c4611G}, we recover that C and O are the main progenitors of Li, Be, and B, but at variance with these results, with the use of the updated cross-section sets, Fe can provide up to $\sim 10\%$ of Li and Be (and $\sim5\%$ of B) at very high rigidity\footnote{These numbers remain consistent with the values reported for Fe in the `Max' column of Tables~V-VII of \citet{2018PhRvC..98c4611G} at 10~GeV/n, which were calculated with \xsGalxxii{}. We stress that the column `Mean' resulted from an average of \xsS{}, \xsW{}, \xsGalxii{}, and \xsGalxxii{} (see their Sect.~\ref{sec:impact}), where 3 out of these 4 parametrisations give a negligible contribution of Fe into LiBeB (see Fig.~\ref{fig:impact_xsupdate} in this paper).}. The remaining contributors are Mg, Ni, Ne, and N, that is the progenitors whose cross sections into LiBeB were updated in this study. We do not detail the multi-step production accounted for in the above plot. For instance, for Li, it amounts to $\lesssim 15\%$ of the total production and peaks at at few GV \citep{2018PhRvC..98c4611G}. Among the important two-step reactions for the production of Li, one finds $^{16}$O$\to^{11}$B$\to^7$Li ($\sim1.5\%$), for which the two associated production cross sections are already updated in our procedure. However, for some others, like the first ranking two-step reaction $^{16}$O$\to^{15}$N$\to^7$Li ($\sim1.9\%$), only one cross section is updated. It is not straightforward to quantify the benefit one would get by renormalising also the relevant cross sections for the most important two-step production channels. It should be a subdominant correction, but updating these cross sections should nevertheless be part of the global strategy to improve the CR flux predictions.

We finally comment on the rigidity dependence of the contributing fractions seen in Fig.~\ref{fig:ranking_per_element}. We recall that the heavier the species, the larger the inelastic cross section, implying a more damped flux at low rigidities (when the interaction rate is larger than the diffusion rate). As a result, ratios of heavier-to-lighter primary CRs, while
asymptotically constant at high energy (we assume the same injection spectrum for all species heavier than C here), decrease with decreasing energy below a few tens of GeV/n; this behaviour was highlighted in Fig.~14 of \citet{2011A&A...526A.101P} and is also seen in the recent AMS-02 Fe/O data (see Fig.~3 of \citealt{2021PhRvL.126d1104A}). Coming back to Fig.~\ref{fig:ranking_per_element}, heavy elements represent a growing fraction of the total CR species with growing energy, and this explains why the contributing fraction of Fe in LiBeB (red dash-dotted curves) grows with energy\footnote{Another subtler energy dependence comes from the fact that $f_{Z_i\to Z_j}$ includes direct and multi-step contributions (from $f_{i\to j}$ coefficients) in variable amounts, with the multi-step production peaking at a few GeV/n \citep{2018PhRvC..98c4611G}.}, with respect to C and O contributions.

\section{Consequences of the updated cross sections: LiBeB/C re-analysis}
\label{sec:LiBeB_reanalysis}

To assess the impact of using the updated production cross-section sets, we re-analyse the AMS-02 LiBeB/C data.  We stress that throughout this section, the cross sections used for the propagation runs are the cumulative ones. These cumulative production cross-section sets are generated from the individual reactions, following Eq.~(\ref{eq:ghost}). We thus have three sets of cumulative cross sections that we keep denoting \optxii{}, \optxiiupxxii{}, and \optxxii{} for simplicity.

\paragraph{Methodology}
We repeat below the analysis of \citet{2020A&A...639A.131W}, which is based on AMS-02 Li/C, Be/C, and B/C data from 2011-2016 \citep{2018PhRvL.120b1101A}\footnote{Data for 7 years were recently presented in \citet{2021PhR...894....1A}, but for the sake of comparison with our earlier analyses, we prefer to stick to the same AMS-02 dataset as used in \citet{2020A&A...639A.131W,2020A&A...639A..74W}.}. In detail, we perform a $\chi^2$ analysis with the \minuit{} minimiser~\citep{1975CoPhC..10..343J}, where $\chi^2$ values are obtained from the sum over (i) the quadratic distance between the model and data, accounting for a covariance matrix ${\cal C}$ of data uncertainties \citep{2019A&A...627A.158D,2019PhRvD..99l3028G,2020A&A...639A.131W}; (ii) a Gaussian-distributed nuisance parameter ${\cal N}_{\phi}$ for the Solar modulation Fisk potential $\phi$ \citep{1967ApJ...149L.115G,1968ApJ...154.1011G}) with mean $\mu_\phi=676$~MV (appropriate for the LiBeB data taking period) and $\sigma_\phi=100$~MV \citep{2020A&A...639A.131W}; (iii) a series of Gaussian-distributed nuisance parameters ${\cal N}^{r}_{\rm XS}$ associated to a selection of LiBeB production reactions \citep{2019A&A...627A.158D,2021JCAP...07..010D,2021PhRvD.103j3016K}; the latter parameters combine a shift in the normalisation and a tilt on the slope of the cross sections, as introduced in \citet{2019A&A...627A.158D}. Formally, this leads to
\begin{equation}
  \chi^2 = (y_{\rm data}-y_{\rm model})\,{\cal C}^{-1}\,(y_{\rm data}-y_{\rm model}) + {\cal N}_\phi + \sum_{r=0}^{n_r}{\cal N}^r_{\rm XS},
  \label{eq:chi2}
\end{equation}
where $r$ runs over the $n_r$ cross-section reactions used as nuisance (see below). It is also useful to isolate the specific contribution of the nuisance parameters to the total $\chi^2$, that is
\begin{equation}
  \chipernui{} \equiv \left({\cal N}_\phi + \sum_{r=0}^{n_r}{\cal N}^{r}_{\rm XS}\right)/(1+n_r),
  \label{eq:chi2nuis}
\end{equation}
which tells us how far away from our initial guess the post-fit values are.
We recall that a good fit of the model to the data typically correspond to $\chi^2_{\rm min}/{\rm dof}\sim 1$ and should have $\chipernui{}\lesssim 1$ (i.e. nuisance parameters within $1\sigma$ of their central values on average).

\paragraph{Post-fit values $\mu_p$ for cross-section nuisance parameters}

The selection of a few most impacting cross sections to use as nuisance are the same as in \citet{2020A&A...639A.131W}\footnote{Specific numbers for the normalisation factors $\mu_{\rm XS}$ and their uncertainties $\sigma_{\rm XS}$ can be found in Table~B.1 of \citet{2020A&A...639A.131W}.} and they only involve reactions on H targets. For the inelastic interactions the most impacting reactions for LiBeB are $^{12}$C+H, $^{7}$Be+H, and $^{11}$B+H. The production reactions are however the most impacting ones: for each element considered in the analysis, we associate a single proxy reaction $p$, related to the production of an isotope (of this element) from the most impacting progenitor. For Li, Be, and B studied here, the associated proxies are $p\mathrm{(Li)=^{16}O+H\to^{6}Li}$, $p\mathrm{(Be)=^{16}O+H\to^{7}Be}$, and $p\mathrm{(C)=^{12}C+H\to^{11}B}$.
This means that in order to obtain a good fit to LiBeB/C, the associated proxy cross-section values $\sigma_p$ were renormalised into $\mu_p\times\sigma_p$, where $\mu_p$ are the post-fit values.

We stress that using a single isotopic cross section as a proxy to account for the uncertainties in an element production is mandatory to avoid degeneracies that would appear, were several isotopic cross sections used for the same element \citep{2019A&A...627A.158D}. However, an undesired side-effect of using a single proxy is that it biases the prediction of isotopic fluxes and ratios. As we want to use the post-fit values of these proxies to draw conclusions about the need for primary source terms for LiBeB, and also to make predictions for future isotopic ratios measured by AMS-02, these proxies must be handled with care.

\paragraph{Contributing fractions $f_r$ for relevant reactions}
We recall that a given reaction $r$ (from a CR progenitor into a CR fragment) only contributes to a fraction $f_r$ of the overall element production (from all possible CR progenitors). For the Li, Be, and B elements, these fractions have been calculated in \citet{2018PhRvC..98c4611G}. We report below, for each proxy $p$ considered, the $f_r$ associated to that proxy (in boldface) and also to the other isotopes (for the element involved)\footnote{We assume here that these contributing fractions are energy-independent, though it is not exactly the case.}, that is
\begin{eqnarray}
f_{r=\mathrm{^{16}O+H\to(^{6}Li,\,^{7}Li)}} &=& ({\bf 15\%},\,12\%),\nonumber\\
f_{r=\mathrm{^{16}O+H\to(^{7}Be,\,^{9}Be,\,^{10}Be)}} &=& ({\bf 19\%},\,6\%,\,1.3\%),
\label{eq:f_p}\\
f_{r=\mathrm{^{12}C+H\to(^{10}B,\,^{11}B)}} &=& (10\%,\,{\bf 33\%}).\nonumber
\end{eqnarray}

\paragraph{From $\mu_p$ to global production correction $\mu_Z^{(p)}$}
When considering a secondary species $Z$, we can calculate a global correction factor to apply for the overall production of the element $Z$ (from any progenitor), as opposed to the single correction $\mu_p$ applied on the proxy $p$ (i.e. single progenitor into a single isotope). We would obtain exactly the same secondary-to-primary elemental ratio (as with the single proxy reaction correction) when applying the correction factor $\mu_Z^{(p)}$ to all individual production cross sections into isotopes of the element $Z$. This global correction factor $\mu_Z^{(p)}$ is obtained from
\begin{equation}
  (\mu_Z^{(p)}-1) = (\mu_p-1)\times f_p\,.\label{eq:TotNormProd}
\end{equation}
The benefit of this procedure is that the global factor $\mu_Z^{(p)}$ now directly tells us by how much the overall production of an element must be modified to match the data: if $\mu_Z^{(p)}>1$ (resp. $<1$), the overall production must be larger (reps. smaller) that what we assumed in the production cross-section set. 
Considering, for instance, B with its proxy $p=\mathrm{^{12}C+H\to^{11}B}$, assuming $\mu_p=1.2$ and applying the above equation would give $\mu_Z^{(p)}=(1.2-1)\times33\%+1=1.067$. This would mean that overall, the production reactions of $B$ are off and must be increased by $6.7\%$ to match AMS-02 LiBeB/C data.
The $\mu_Z^{(p)}$ values are used for the discussion about evidences for LiBeB primary source terms in Sect.~\ref{sec:LiSource} (see Fig.~\ref{fig:ellipses_slim}).

\paragraph{Unbiased isotopic production cross sections via $\mu_{\rm iso}$}
Applying a correction on the proxy reaction only (e.g. $\mu_{p=\mathrm{^{12}C+H\to^{11}B}}=1.2$ in the above example) changes the predictions for CR isotopic ratios without real justification. To avoid this, one possibility is to redistribute this modification democratically between the isotopes (e.g. $^{10}$B and $^{11}$B here). In practice, we build new production cross-section sets (from our updated sets), applying the correction factor $\mu_Z^{{\rm iso}}$ for all reactions from the progenitor (used in the proxy) into the isotopes of the element (to which the proxy is associated), with
\begin{equation}
  (\mu_Z^{{\rm iso}}-1) \times \sum_{i \in Z} f_i =  (\mu_p -1)\times f_p\,.
  \label{eq:mu_D}
\end{equation}
For instance, the proxy $p=\mathrm{^{12}C+H\to^{11}B}$ involves the progenitor $^{12}$C, so that using the above equation, we apply the correction $\mu_Z^{{\rm iso}}=[(1.2-1)\times33\%]/43\%+1=1.155$\footnote{We calculated $\sum_{i\in {\rm B}} f_i= f_{\mathrm{^{12}C+H\to^{10}B}}+f_{\mathrm{^{12}C+H\to^{11}B}} = 43\%$.} to both the cross sections $\sigma_{^{12}{\rm C+H}\to^{10}{\rm B}}$ and $\sigma_{^{12}{\rm C+H}\to^{11}{\rm B}}$ (instead of using the single proxy correction  $\mu_p\times\sigma_{\mathrm{^{12}C+H\to^{11}B}}$). These new sets are identified by the suffix `-U' (for unbiased)---that is we build \optxii{}-U, \optxiiupxxii{}-U, and \optxxii{}-U---and are used for isotopic ratio calculations in Sect.~\ref{sec:isot_ratios} (and Fig.~\ref{fig:isotopic_ratios_democratic}); and we explicitly checked that these sets leave unchanged the predictions for the elemental fluxes (as expected).

Now that we are done with the definitions, we can move on to the interpretation of our LiBeB/C re-analysis results: first, we discuss the impact of the new cross-section sets \optxii{}, \optxiiupxxii{}, and \optxxii{} on the transport parameters (Sect.~\ref{sec:impact_transport}); second, we address the need for a primary lithium source (Sect.~\ref{sec:LiSource}); finally, we study the impact of the updated cross-section sets on CR isotopic ratios (Sect.~\ref{sec:isot_ratios}).

\subsection{Impact on transport parameters}
\label{sec:impact_transport}
\begin{figure}[t!]
   \centering
   \includegraphics[width=\columnwidth]{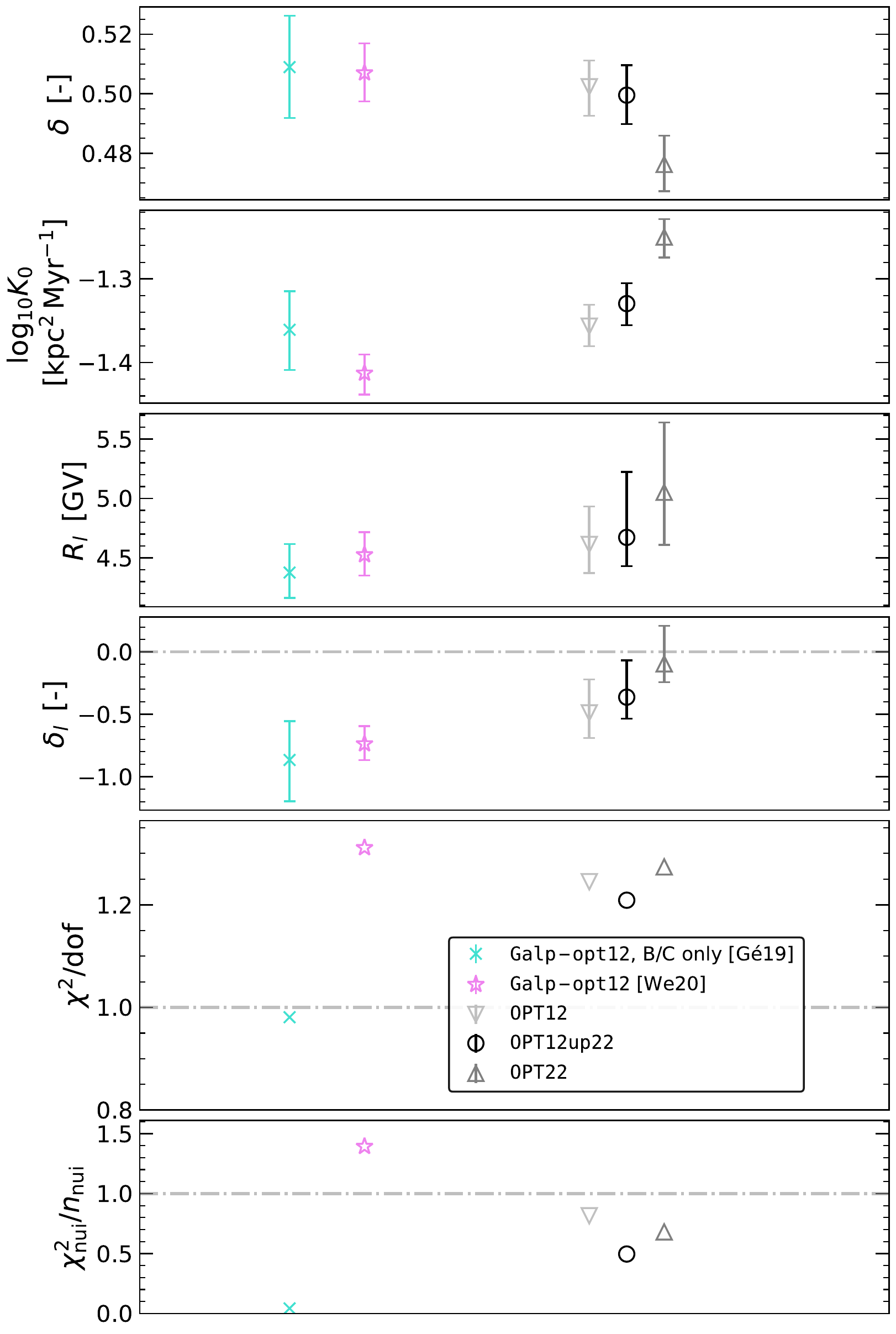}
   \caption{Best-fit transport parameters (and uncertainties) from the fit to AMS-02 LiBeB/C data using different cross-section sets: from left to right, the original \xsGalxii{} set used in \citet{2019PhRvD..99l3028G} and \citet{2020A&A...639A.131W}, and the three updated sets obtained in this paper (\optxii{}, \optxiiupxxii{}, and \optxxii{}). The next-to-last and last panels show the \chimindof{} and \chipernui{} values for the associated runs---see Eqs~(\ref{eq:chi2}) and (\ref{eq:chi2nuis}) respectively.
   \label{fig:slim_pars}}
\end{figure}

In this subsection, we show the constraints set on the \SLIM{} parameters at $L=5$~kpc (see App.~\ref{app:update_transport} for the constraints on \BIG{} and \QUAINT{}). The minimisations involve 7 nuisance parameters (one for Solar modulation and 6 for cross sections) and 201 LiBeB/C data points. We show in Fig.~\ref{fig:slim_pars} the best-fit transport parameters of \SLIM{}, that is, from top to bottom, the normalisation and slope of the diffusion coefficient ($K_0$ and $\delta$), and the low-rigidity break parameters ($\delta_l$ and $R_l$, see Eq.~\ref{eq:def_K}) for different cross-section sets; the next-to last and last panels show $\chi^2_{\rm min}/{\rm dof}$ and $\chipernui{}$ respectively, see Eqs.~(\ref{eq:chi2}) and (\ref{eq:chi2nuis}). We show on the right-hand side of the plot the values obtained in this analysis with the three updated cross-section sets \optxii{}, \optxiiupxxii{}, and \optxxii{}. For comparison purpose, we also report, on the left-hand side of the plot, the parameters obtained in \citet{2020A&A...639A.131W} from the same LiBeB analysis but with the original set \xsGalxii{} (violet stars), and those obtained even earlier from the same cross-section set in \citet{2019PhRvD..99l3028G} where only a fit to AMS-02 B/C data was performed\footnote{The original analysis was based on $L=10$~kpc and the values for $L=5$~kpc are obtained using the scaling relations of \citet{2020A&A...639A..74W}.} (cyan crosses). The latter result is shown to recall that, with respect to a combined analysis of several secondary-to-primary species, the analysis of a single ratio is both less constraining (larger error bars) and possibly biased by systematics in the production cross sections \citep{2019A&A...627A.158D,2020A&A...639A.131W}.

\begin{figure}[t!]
   \centering
   \includegraphics[width=\columnwidth]{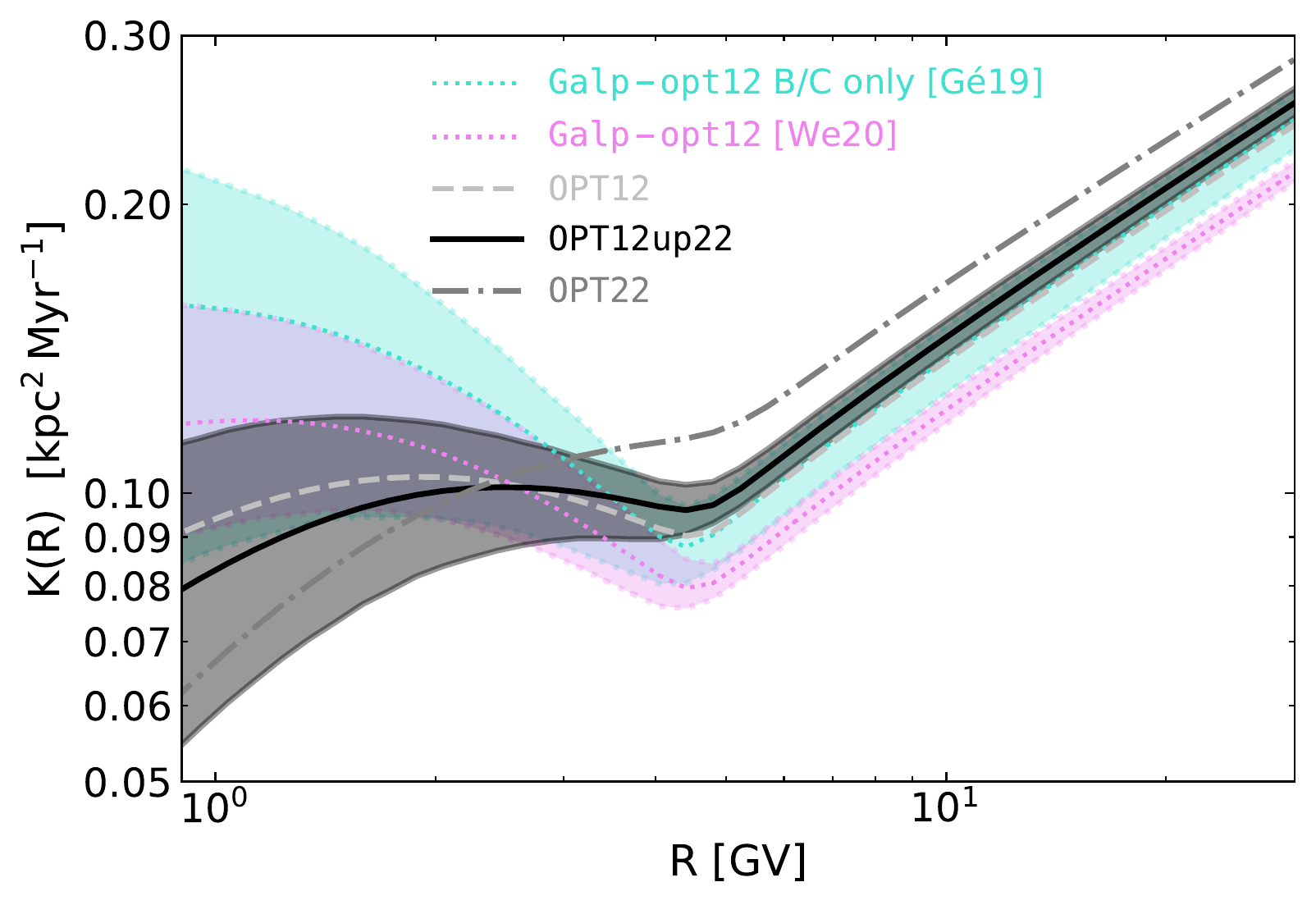}
   \caption{Best-fit and 1$\sigma$ envelopes for the diffusion coefficient, Eq.~(\ref{eq:def_K}), associated to the parameters shown in Fig.~\ref{fig:slim_pars} and for the same cross-section sets ($L=5$~kpc). For readability, we do not show the envelopes for \optxii{} and \optxxii{} as they are similar to those for \optxiiupxxii{}.
   \label{fig:KR}}
\end{figure}

The striking features seen in Fig.~\ref{fig:slim_pars} for the combined LiBeB/C fit are the following. Firstly, when going from \optxii{} to \optxxii{}, the normalisation $K_0$ (second panel) grows, whereas the low-rigidity break $\delta_l$ becomes consistent with $0$ (i.e. no break). The change in $K_0$ is directly linked to the updated cross sections, as more production means less grammage to produce the same amount of secondary species, which in turns means a larger diffusion coefficient (at fixed halo size $L$). The overall impact on the diffusion coefficient $K(R)$ is also illustrated in Fig.~\ref{fig:KR} where $1\sigma$ envelopes are shown.
Secondly, not only do the updated cross-section sets improve the global fit substantially, but they also reduce the deviations of the nuisance parameters from their priors. In particular, the best \chidof{} is obtained for \optxiiupxxii{}\footnote{We recall that the fact that \chidof{} is significantly larger than 1 is caused by the two low-rigidity Be/B points (upturn) that cannot be well-fitted by the models, see \citet{2020A&A...639A.131W}.} (black circle), with a very significant improvement ($\chi^2_{\rm min}$ goes from $258.3$ to $237.5$) compared to the use of the original \xsGalxii{} set in \citet{2020A&A...639A.131W} analysis (violet star). This \optxiiupxxii{} set also gives the smallest \chipernui{} value, making it the most favoured by the LiBeB/C data.

\begin{figure}[t!]
   \centering
   \includegraphics[width=\columnwidth]{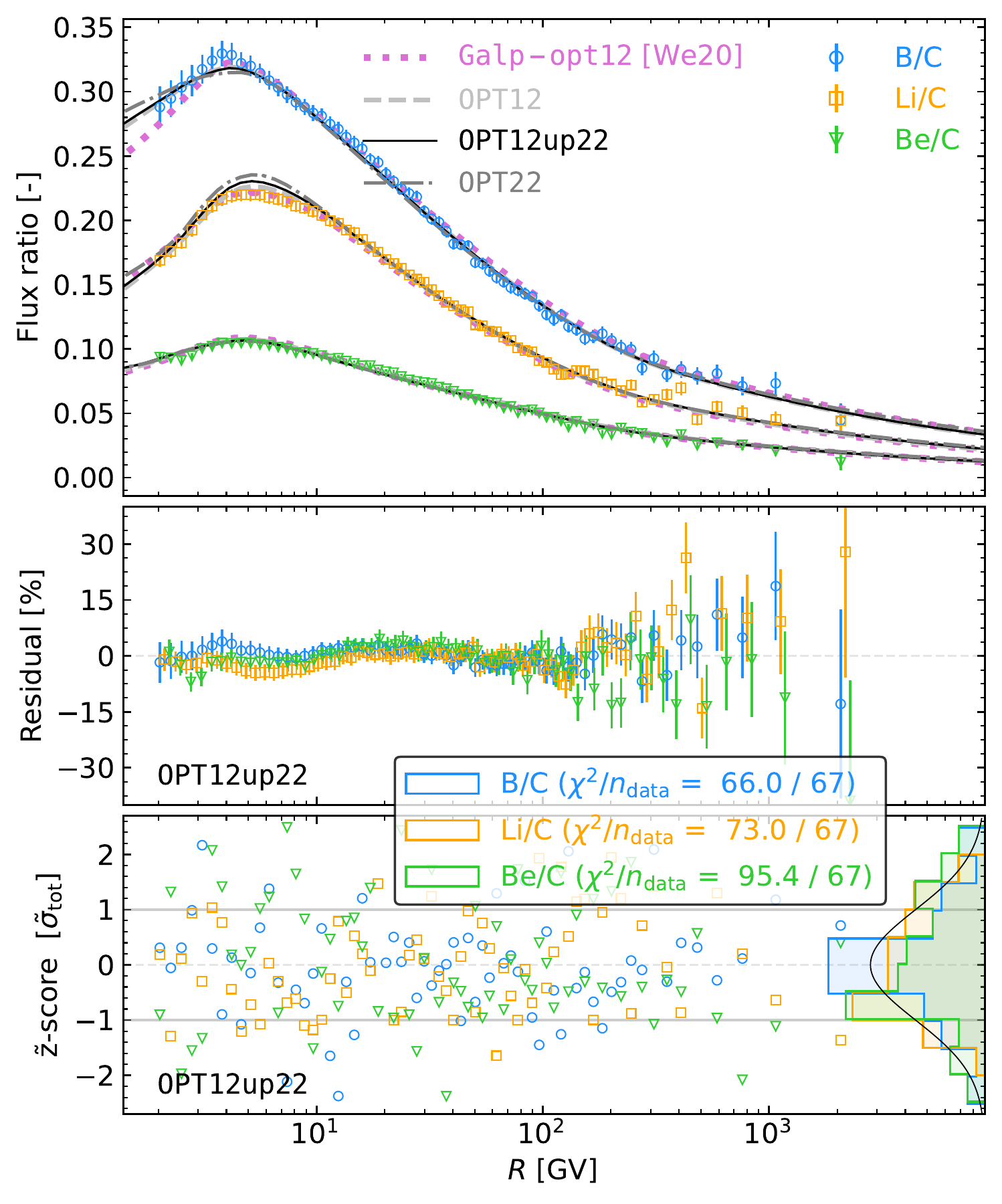}
   \caption{Flux ratios (top), residuals (centre) and $\tilde{z}$-scores (bottom) for B/C (blue circles), Be/C (green triangles), and Li/C (orange squares). The models (top panel) are calculated from the best-fit transport parameters (combined LiBeB/C analysis) using the original \xsGalxii{} set (thick violet dotted line) or our updated sets: \optxii{} (dashed grey line), \optxiiupxxii{} (solid black line), and \optxxii{} (dashed-dotted grey line). The residuals and $\tilde{z}$-score are shown for \SLIM{} only---the $\tilde{z}$-score is related to the usual $z$-score by a rotation in a base where the covariance matrix of data systematics is diagonal (see text for details).
   \label{fig:LiBeB_model_vs_data}}
\end{figure}

For illustration purpose, the best-fit models for Li/C, Be/B, and B/C ratios are reported (\SLIM{} configuration for the various cross-section sets) and compared to the data in the top panel of Fig.~\ref{fig:LiBeB_model_vs_data}.
For the specific \optxiiupxxii{} cross-section set, we further show in the middle and bottom panels respectively, the residuals with regard to the total uncertainties (systematics and statistic combined) and the rotated score $\tilde{z}$; the latter quantity accounts for the role of correlations in the data systematics\footnote{As introduced in \citet{2020PhRvR...2b3022B}, a graphical representation of the `rotated' score (denoted $\tilde{z}$-score) is better suited to provide an unbiased graphical view of the difference between the model and data, when correlations exists between the data bins. The rotated base is defined so that the covariance matrix of uncertainties is diagonal, $\tilde{\cal C} = U {\cal C}U^{\rm T}$, with $U$ an orthogonal rotation matrix. In the rotated base, rotated rigidities are $\tilde{R}_{i} = \sum_{j} U_{ij}^{2} \, R_{j}$ (that remain close to the original $R_{i}$ values for AMS-02 data), and the rotated residuals become $\tilde{z}_{i} = \tilde{x}_{i}/\sqrt{{\cal C}_{ii}}$ with $\tilde{x}_{i} \equiv \sum_j U_{ij} ({\rm model}_j-{\rm data}_j)$. By construction, $\chi^2 = \sum_{i} \tilde{z}_{i}^{2}$ (distance between the model and data, forgetting about the nuisance parameters), and we further stress that for a $\chi^2/n_{\rm data}$ close to 1, the histogram of $\tilde{z}_{i}$ values should follow a centred Gaussian of width unity (see right-hand side of the bottom panel in Fig.~\ref{fig:LiBeB_model_vs_data}).}.
Overall, the above plot is very similar to the corresponding one shown in Fig.~2 of \citet{2020A&A...639A.131W}, that is the model matches the data quite well. Moreover, very similar $\chi^2/n_{\rm data}$ values (as reported in the legend of the second panel) and goodness-of-fit are obtained for the \BIG{} and \QUAINT{} configurations (not shown).

\subsection{Is a primary source of Li needed}
\label{sec:LiSource}

\begin{figure}[t!]
   \centering
   \includegraphics[width=\columnwidth]{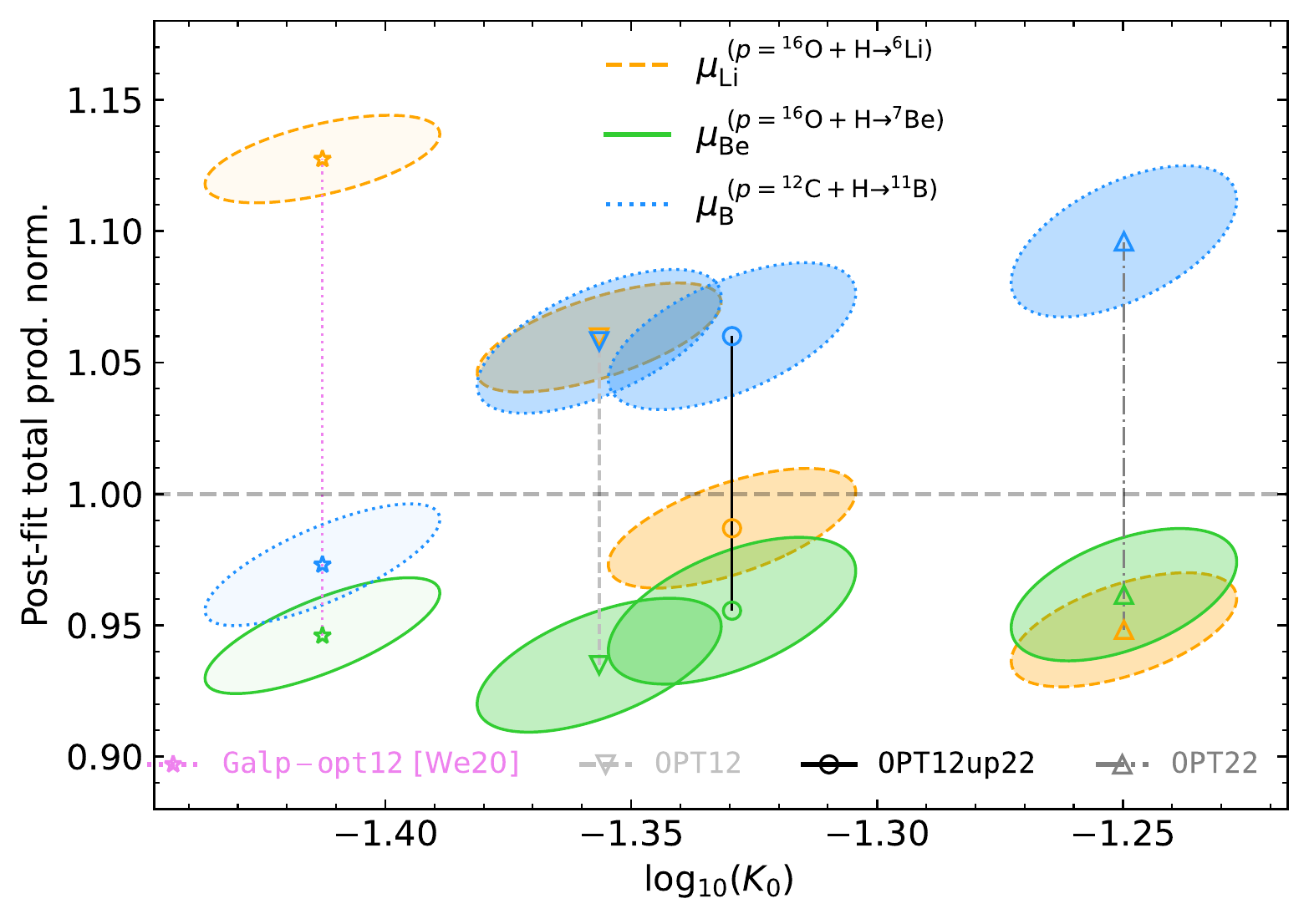}
   \caption{Correlation between log$_{10}(K_{0})$ and the normalisation factor $\mu_Z^{(p)}$ (see Eq.~\ref{eq:TotNormProd}), that is the correction factor applied (on the total production cross section of element $Z$) in order to get the best-fit on LiBeB/C data. The three elements considered are colour-coded: $Z=3$ (Li) in orange, $Z=4$ (Be) in green, and $Z=5$ (B) in blue. The $1\sigma$ correlation ellipses are shown for analyses with different cross section sets (in model \SLIM{}): from left to right, stars are for the original \optxii{}, downward triangles for the update \optxii{}, circles for \optxiiupxxii{}, and upward triangles for \optxxii{}.
   The horizontal grey dashed line highlights $\mu_Z=1$ (i.e. no modification needed for the production cross sections).
   \label{fig:ellipses_slim}}
\end{figure}
We recall that in \citet{2020ApJ...889..167B}, using the propagation code \galprop{} and no nuisance parameters on the cross sections, the authors proposed the presence of a primary Li contribution to correct for a 20-25\% deficit between the model and the data. Conversely, using nuisance parameters and relying on the same cross-section set \xsGalxii{}, \citet{2020A&A...639A.131W} found no mismatch between the model and the data, but rather the need for a 13\% increase of the production cross section of Li (see also \citealt{2021JCAP...03..099D,2021JCAP...07..010D} for similar conclusions, though with slightly different numbers). The latter conclusion was reached by inspecting $\mu_Z^{(p)}$ values calculated from Eq.~(\ref{eq:TotNormProd}), reproduced here with star symbols (and their associated ellipses) in the left-hand side of Fig.~\ref{fig:ellipses_slim}: in \citet{2020A&A...639A.131W}'s analysis, to match the LiBeB/C data with the  \xsGalxii{} cross-section set, the overall production of Li had to be increased by $13\%$, and that for B (resp. Be) had to be decreased by 3\% (resp. 6\%).

This picture changes when considering our updated cross-section sets: moving from the original \xsGalxii{} to the updated sets (from left to right in the plot), the total Li production goes from a significant increase ($+13\%$) to a mild decrease ($-5\%$), whereas the total B production evolves in the opposite direction (from $-3\%$ to $+10\%$); the overall production of Be remains stable at $\sim-5\%$. This shows that there is no need for a primary source of Li, and that we cannot trust the fact that, taken at face value, we would need now a primary source of B. Our analysis merely reinforces the conclusions of \citet{2020A&A...639A.131W} and \citet{2021JCAP...03..099D}: AMS-02 LiBeB/C data can be reproduced without invoking the presence of a new primary source because mismatches are likely to be associated to cross-section uncertainties.
We also note in passing that \optxiiupxxii{}, which was our (slightly) favoured configuration from the nuclear data perspective (see Sect.~\ref{sec:xs_rescaling}), is the one that best-fit AMS-02 LiBeB/C data (see Fig.~\ref{fig:slim_pars}). It is also the one for which $\mu_Z^{(p)}$ values are the closest to 1 (within $\pm5\%$, see empty circles in Fig.~\ref{fig:ellipses_slim}), that is minimal modifications needed for the production cross sections.

Figure~\ref{fig:ellipses_slim} and the discussion are based on the \SLIM{} configuration, but we checked that the \BIG{} and \QUAINT{} configurations lead to the same decrease/increase of the Li, Be, and B production cross sections (and their uncertainties), that is same values as those shown in Fig.~\ref{fig:ellipses_slim} along the $y$ axis. Our conclusion against the evidence of a Li primary source in CRs is thus independent of the propagation configuration and robust.

\subsection{Impact on isotopic ratios}
\label{sec:isot_ratios}

It is also interesting to see whether the updated cross-section sets predict different isotopic contents for the LiBeB elements. The HELIX project \citep{2019ICRC...36..121P} and the AMS-02 \citep{2019PhRvL.123r1102A} experiment both have the capability to separate light isotopes from a few hundreds of MeV/n up to $\sim 10$~GeV/n , whereas past experiments only provide data on a much limited range; the most recent data on $Z=3-5$ isotopic fluxes are from ACE \citep{2006AdSpR..38.1558D}, AMS-01 \citep{2011ApJ...736..105A}, ISOMAX \citep{2004ApJ...611..892H}, and PAMELA \citep{2018ApJ...862..141M,2021Univ....7..183N}.

\begin{figure}[t!]
   \centering{}
   \includegraphics[width=\columnwidth]{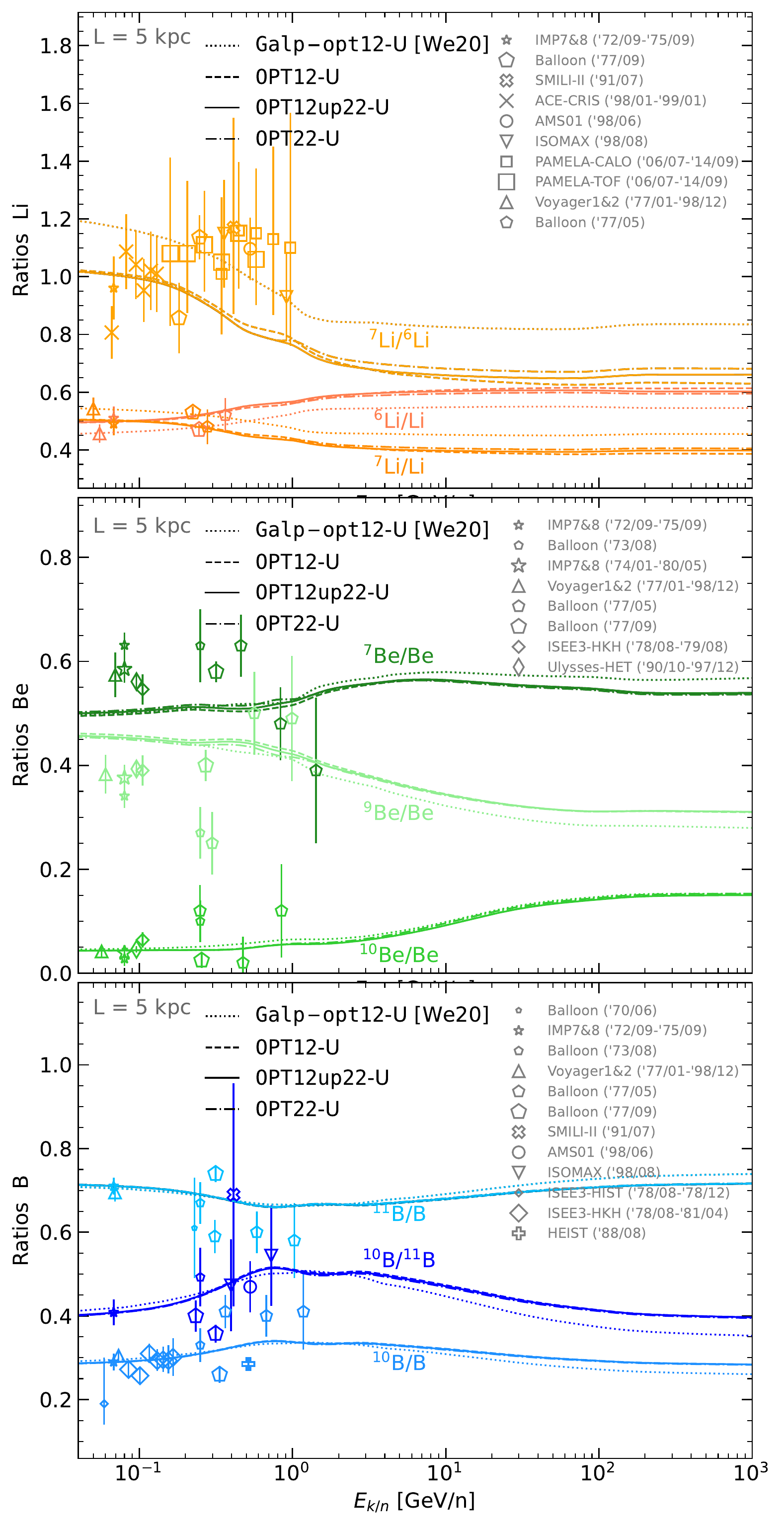}
   \caption{Isotopic ratios of Li, Be, and B (from top to  bottom panel) calculated from the best-fit models to AMS-02 LiBeB/C (at fixed $L=5$~kpc). The production cross-section sets considered are the original \xsGalxii{} (dotted lines) and the updated sets \optxii{} (dashed lines), \optxiiupxxii{} (solid lines), and \optxxii{} (dash-dotted lines); the suffix `-U' in the labels recalls that we use here unbiased cross-section sets, see Eq.~(\ref{eq:mu_D}). The data are from IMP \citep{1977ICRC....1..301G,1977ApJ...217..859G,1981ICRC....2...72G}, Balloon '73/08 \citep{1977ApJ...212..262H}, Balloon '77/05 \citep{1978ApJ...226..355B}, Balloon '77/09 \citep{1979ICRC....1..389W}, Balloon '70/06 \citep{1979ZPhyA.291..383B}, ISEE3 \citep{1980ApJ...239L.139W,1981ApJ...251L..27M,1988ApJ...328..940K}, HEIST \citep{1992ApJ...391L..89G}, Ulysses \citep{1998ApJ...501L..59C}, Voyager1\&2 \citep{1999ICRC....3...41L}, SMILI-II \citep{2000ApJ...534..757A}, ISOMAX \citep{2004ApJ...611..892H}, ACE \citep{2001ApJ...563..768Y,2006AdSpR..38.1558D}, AMS-01 \citep{2011ApJ...736..105A}, PAMELA \citep{2018ApJ...862..141M,2021Univ....7..183N}.
   \label{fig:isotopic_ratios_democratic}
   }
\end{figure}
We show in Fig.~\ref{fig:isotopic_ratios_democratic} various isotopic ratios of Li, Be, and B, calculated from the best-fit models (parameters shown in Sect.~\ref{sec:impact_transport}) for various `unbiased' production cross-section sets. We see that the updated sets predict very similar isotopic ratios for all species, except for Li isotopes, for which the largest difference is seen for $^7$Li/$^6$Li; this is not so surprising since all sets were renormalised to the same nuclear data (though the underlying original cross-section models have different energy dependences, and possibly different prediction for unmeasured reactions). The largest difference for $^7$Li/$^6$Li is at high energy and disappears at low energy. It is likely to be associated with Fe fragmentation: as seen in Fig.~\ref{fig:ranking_per_element}, the contribution from Fe to the total production vanishes at low energy, and as seen on the top left panel of Fig.~\ref{fig:xs_data_vs_model_2}, this is where the updated cross-section values differ most for both $^6$Li and $^7$ Li. The difference between \xsGalxii{} (dashed lines) and the updated cross-section sets (all other lines) is also maximal for Li isotopes: the origin of the difference can be read off Table~\ref{tab:LiBeB_XS_rescalingfactor_HE}, where one can see that the production cross sections from $^{16}$O and $^{12}$C have been increased for $^6$Li and decreased for $^7$Li. The differences seen for the other isotopes are smaller, but also result from the updated cross sections listed in Table~\ref{tab:LiBeB_XS_rescalingfactor_HE}. This behaviour is qualitatively (but not quantitatively) the same if we consider the \BIG{} and \QUAINT{} transport configurations instead of \SLIM{}.

We finally want to stress that the suffix `-U' in the labels of Fig.~\ref{fig:isotopic_ratios_democratic}) indicate that our predictions were made with the `unbiased' cross-section sets. In these sets, the corrections applied to cross-sections for a given element are distributed equally between the various isotopes of this element---parameter $\mu_{\rm iso}$ in Eq.~(\ref{eq:mu_D}). We recognise that this procedure is debatable. Indeed, once the fitting procedure to secondary elements decides that the overall production of the said element must be slightly increased (or decreased), there is no real clear path to choose how to distribute this modification over the various isotopes (i.e. modification of a specific isotope or equally distributed between isotopes). Using more proxy reactions (e.g. one proxy per isotope) must be backed-up by isotopic ratio data in the fit, otherwise the various proxy parameters are degenerated; in that latter case, post-fit value would have no predictive power and would lead to overestimated variations in the predicted isotopic ratios.
As illustrated in Fig.~\ref{fig:isotopic_ratios_democratic}, current data (symbols) are clearly too scattered and imprecise to get a stronger handle on the isotopic production cross sections. Moreover, at this stage, it is difficult to quantify whether our updated (and `unbiased') cross-section sets provide an improvement over the use of the original \xsGalxii{}-U set. The forthcoming AMS-02 and HELIX data are awaited to make some progress on these issues.

\section{Conclusions}
\label{sec:conclusions}

We have revisited and updated the production cross sections of Li, Be, and B species. We showed that contributions from Fe favour an enhanced production of Li---and to a lesser extent Be---with respect to B. We emphasised that most (and possibly all) previous interpretations of the AMS-02 Li, Be, and B data rely on a cross-section set that does not account for these enhancements. The consequences of using updated cross sections are the following\footnote{These cross-section sets will appear at some point in the next release of the \usine{} code. They can be obtained on demand in the meantime.}:
(i) in the ranking of the most important progenitors for LiBeB, Fe becomes the third most important element, responsible for $\sim10\%$ of Li and Be\footnote{In practice, this slightly modifies the priority order (for new measurement campaigns in beam+target experiments) discussed in Fig.~5 of \citet{2018PhRvC..98c4611G}.}, but only $\sim4\%$ for B; 
(ii) the combined analysis of the AMS-02 Li/C, Be/C, and B/C ratios lead to a 20\% larger value of the diffusion coefficient (compared to those obtained in our previous investigation), which is not without consequences for the predictions made, for instance, on secondary antiprotons \citep{2020PhRvR...2b3022B}\footnote{It means that this is another worrying systematics to consider when trying to uncover or set limits on dark matter in this channel \citep[e.g.][]{2022ScPP...12..163C}.};
(iii) the updated cross-section sets also impact the predictions for the isotopic ratios of Li, Be, and B---in particular, uncertainties on the production of Be isotopes may be an issue for the determination of the halo size of the Galaxy (see the companion paper, \citealt{2022arXiv220307265M});
(iv) via the inspection of the post-fit values of the nuisance parameters (on the production cross section), we find that the updated cross-section sets, expected to be closer to the truth, indeed provide a more consistent description of the Li/C, Be/C, and B/C data---this is reassuring and validates to some extent our approach.

We recall that the motivation for this re-analysis was the interpretation of a mismatch between models and AMS-02 Li data, interpreted by some authors as a possible hint for a primary source of Li. We unambiguously showed here that the need for an extra Li component completely disappears, comforting the studies arguing for systematics in the production cross sections instead; at variance with these studies, however, we showed that these systematics were mostly attributed to an underestimated production of Li from Fe. Actually, the different plausible cross-section sets we built either lead to the need for a $5-10\%$ increase or decrease of Li and B production cross sections; our favoured cross-section set (that also happens to provide the best-fit to Li/C, Be/B, and B/C data) even leads to a very small excess of Li with respect to the data. For this reason, we caution against any hasty interpretation of these overshoots or undershoots: they merely reflect the presence of systematics or still missing important reactions in the production cross-section sets.

The next step of this work would be to generalise our normalisation procedure (on nuclear data) to sub-dominant production channels. It could also be useful to make systematic comparisons to state-of-the-art cross-section codes like SPACS \citep{2014PhRvC..90f4605S}, FRACS \citep{2017PhRvC..95c4608M}, or others, in order to better quantify systematics related to unmeasured cross sections. The rescaling procedure should then be extended to the production of  secondary species with $Z>5$, as all the elements up to Fe are within reach of the AMS-02 experiment.
On the nuclear data side, new measurements for the production of Li and B isotopes from C and O, but also from Ne, Mg, Si, and Fe progenitors would be of great value: these progenitors amount to $\sim 20\%$ of the total production of Li and B, while many of these reactions remain unmeasured or with only a couple of data points.

\begin{acknowledgements}
D.M. thanks his AMS-02 colleagues for interesting discussions that lead to the implementation of a ranking per progenitor shown in Fig.~\ref{fig:ranking_per_element}.
We thank the Center for Information Technology of the University of Groningen for their support and for providing access to the Peregrine high-performance computing cluster.
This work was supported by the Programme National des Hautes Energies of CNRS/INSU with INP and IN2P3, co-funded by CEA and CNES. Y.G. acknowledges support from \textsc{Villum Fonden} under project no.~18994.
\end{acknowledgements}

\appendix

\section{Impact of updated reactions}
\label{app:impact_per_progenitor}

We showed in the main text the global impact of the updated cross sections on the calculated Li/C, Be/C, and B/C ratios (Fig.~\ref{fig:xs_old_vs_new}). We show here a more detailed view, identifying the progenitors whose updated cross sections make up most of the observed changes in these ratios.
However, we do not comment on the energy dependence seen in the figures, as they are already discussed in the main text (see Sect.~\ref{sec:xs_impact}).

\begin{figure}[t!]
   \centering
   \includegraphics[width=0.5\textwidth]{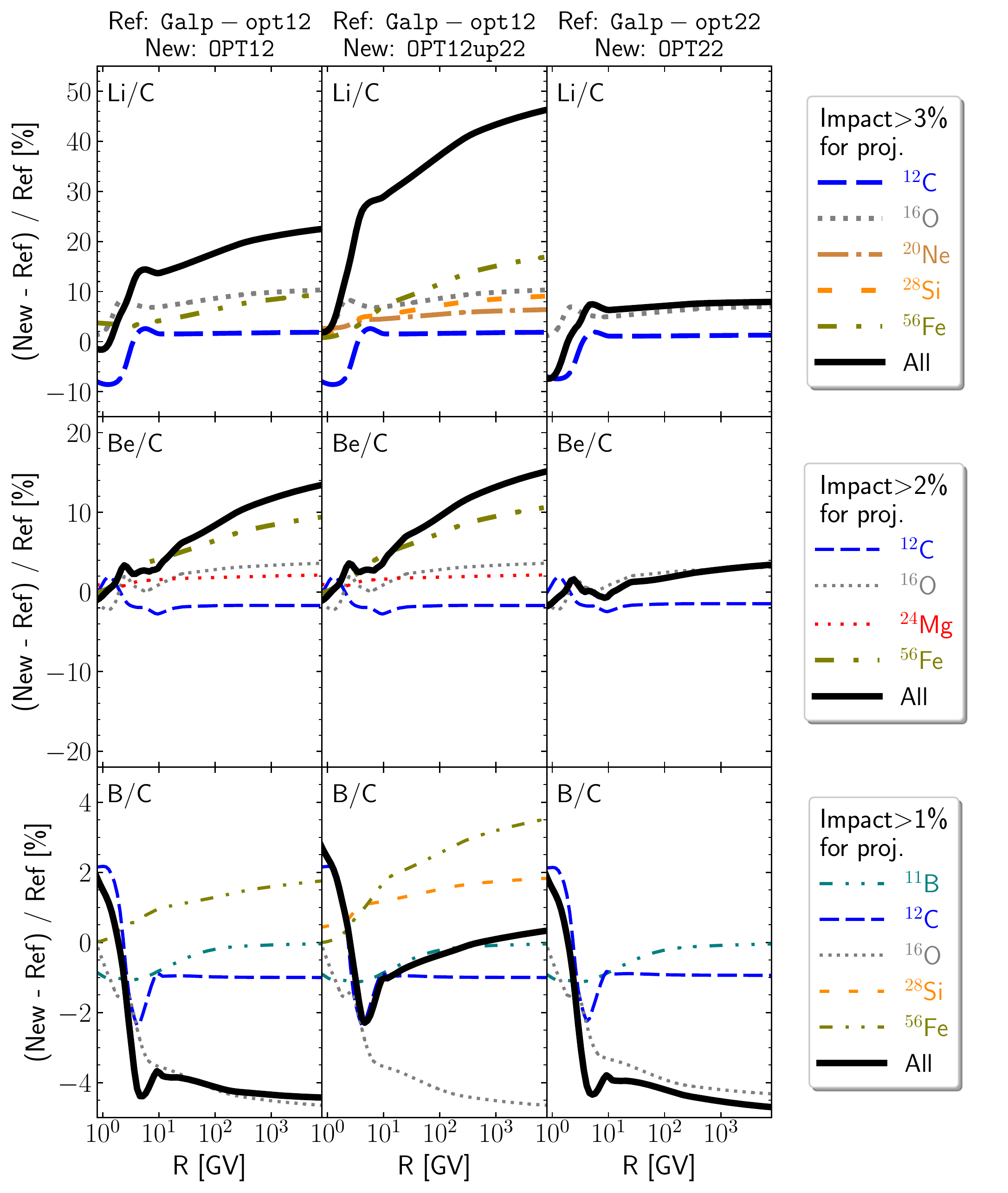}
   \caption{Global impact (thick black lines) of updated cross-sections in Li/C (top panels), Be/C (centre panels), and B/C (bottom panels) against broken-down impact per progenitor (colour-coded), as a function of rigidity. Rows show only progenitors whose relative impact is larger than 3\% (resp. $2\%$ and $1\%)$ for Li/C (resp. Be/C and B/C); all calculations are at a fiducial Solar modulation level of 700~MV. The columns show different ways to update the cross-section sets: updating the original \xsGalxii{} into \optxii{} (left panels); updating the original \xsGalxxii{} into \optxxii{} (right panels), or the hybrid approach \optxiiupxxii{} (centre panels). See text for details.
   \label{fig:impact_xsrescaled_ratios}}
\end{figure}
In the left panels of Fig.~\ref{fig:impact_xsrescaled_ratios}, the reference calculation relies on the original \xsGalxii{} cross-section set, whereas the new calculation relies on the associated updated set \optxii{}. The maximal impact from updating the cross sections (thick solid lines) typically reaches $+20\%$ on Li/C (top), $+10\%$ on Be/C (centre), and $ -5\%$ on B/C (bottom). In all three ratios, the culprits for these changes are the fragmentation of $^{56}$Fe (green double-dashed line) but also of $^{16}$O. This is in line with the values observed for the rescaling factors ${\cal R}$ shown in Table~\ref{tab:LiBeB_XS_rescalingfactor_HE}, where ${\cal R}$ varies the most for the $^{56}$Fe and $^{16}$O progenitors.

In the right panels, the reference calculation is now the original \xsGalxxii{} set, whereas the new calculation is based on the associated updated set \optxxii{}. In that case, updating $^{56}$Fe reactions has a more limited impact on the flux ratios. This is consistent with the observation that the initial cross-section set \xsGalxxii{} was slightly overshooting the nuclear data: in Table~\ref{tab:LiBeB_XS_rescalingfactor_HE}, the rescaling for $^{56}$Fe (second number in the entry associated to \optxxii{}) is, on average, close to 1 (i.e. the model prediction was already close to the data). In addition, with Fe fragmentation ranking at the $\lesssim 10\%$ level (see Fig.~\ref{fig:ranking_per_element}), the combination of these last two numbers falls in the percent range at most. In this context, the most significant impact in LiBeB/C is from updating $^{16}$O fragmentation reactions.

Finally, the middle panels shows the hybrid case, where the initial parametrisation is \xsGalxii{} (for all measured and unmeasured cross sections), but where we substitute for measured cross section an update of the original \xsGalxxii{} (so called \optxiiupxxii{} set); we recall that this was motivated by the more realistic energy dependence provided in the \xsGalxxii{}. This configuration is expected to be the most realistic, and we see that the ratios are maximally impacted, with Li/C maximally enhanced (up to $\sim 40\%$). Moreover, these changes are now dominated by the updated Fe fragmentation.

\section{Impact of recent F, Ne, Na, Mg, Al, Si, and Fe AMS-02 data}
\label{app:impact_AMS}

In a propagation run, while we do not fit the primary species, we rescale primary source abundances on a specific high-energy point. As illustrated in \citet{2019PhRvD..99l3028G}, this procedure is sufficient to match primary CR progenitors of LiBeB (provided we fix correctly the universal source slope). Moreover, the secondary-to-primary ratios are insensitive to this source slope \citep{2002A&A...394.1039M,2015A&A...580A...9G}.
\begin{figure}[t!]
  \includegraphics[width=\columnwidth]{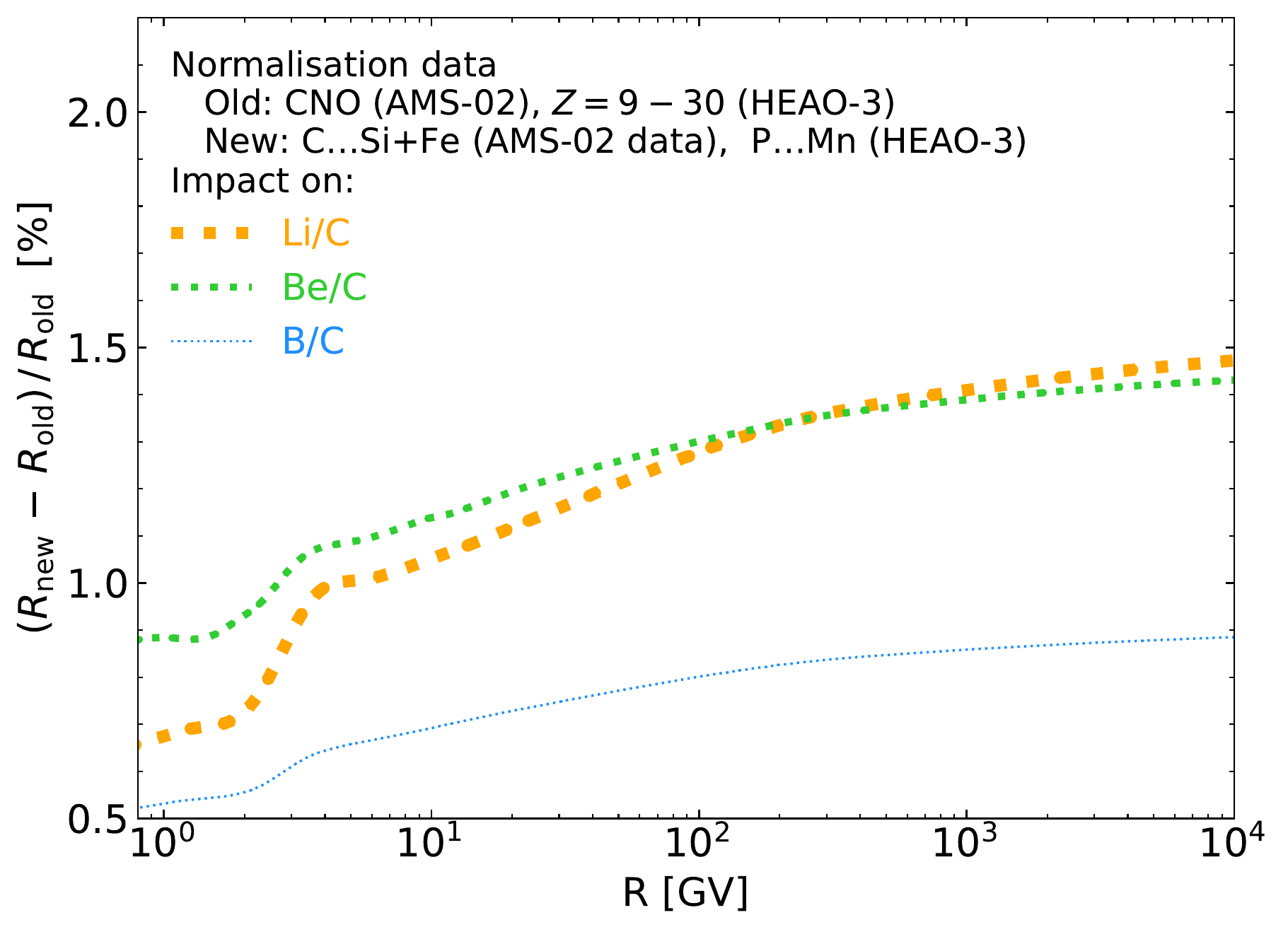}
  \caption{Impact of using the recently published AMS-02 data in the LiBeB/C calculation. The calculation relies on the use of C, N, O, F, Ne, Na, Mg, Al, Si, and Fe data from AMS-02, and HEAO-3 data for all other elements, whereas our previous analysis \citep{2020A&A...639A.131W},  relied on C, N, O AMS-02 data (only available LiBeB progenitors from this experiment at the time) and HEAO-3 data for all other progenitors.
  \label{fig:impact_norm}}
\end{figure}
In practice, to ensure the most accurate calculation, it is best to use the most-recent data available to normalise these source abundances. We use AMS-02 elemental data at 50~GV when available and HEAO3 data \citep{1990A&A...233...96E} at 10.6 GeV/n otherwise.

In our previous LiBeB/C analysis \citep{2020A&A...639A.131W}, only AMS-02 data for the progenitors C, N, and O were available \citep{2017PhRvL.119y1101A,2018PhRvL.121e1103A}. For this publication, we can rely on AMS-02 data for more progenitors, namely F, Ne, Na, Mg, Al, Si, and Fe \citep{PhysRevLett.126.081102,2020PhRvL.124u1102A,2021PhRvL.126d1104A}, but also C, N, and O fluxes for 7 years of data \citep{2021PhR...894....1A}.
Figure~\ref{fig:impact_norm} illustrates by how much the Li/C (thick yellow dashed line), Be/C (green dashed line), and B/C (thin blue dashed line) ratios are impacted when these new data are taken into account in the calculation. We see that Li/C and Be/C are the most impacted ones, but only at a few percent level. This is understood as follows: first, N (resp. Na and Ne) contributes to the percent level (resp. negligibly) to LiBeB production \citep{2018PhRvC..98c4611G}, so that a small discrepancy between AMS-02 and HEAO3 data would have no impact on the calculated LiBeB; second, Fe is responsible for up to tens of percent of Li, a few percent on Be, and almost no impact on B. This ordering is consistent with the one seen in Fig.~\ref{fig:impact_norm}, and with the fact that AMS-02 data for Fe are slightly above HEAO3 ones \citep{2021PhRvL.126d1104A}.

CALET has recently published new data for the C, O and Fe fluxes \citep{2020PhRvL.125y1102A,2021PhRvL.126x1101A}, which were found to be respectively $\sim 30\%$ and $\sim27\%$ lower than the AMS-02 ones. This difference for C and O would certainly impact the production of lighter elements, but taking ratios of elements, as we do here, cancels out these yet to be explained differences.

\section{Further checks on transport parameters}
\label{app:update_transport}

\begin{figure}[t!]
   \centering
   \includegraphics[width=\columnwidth]{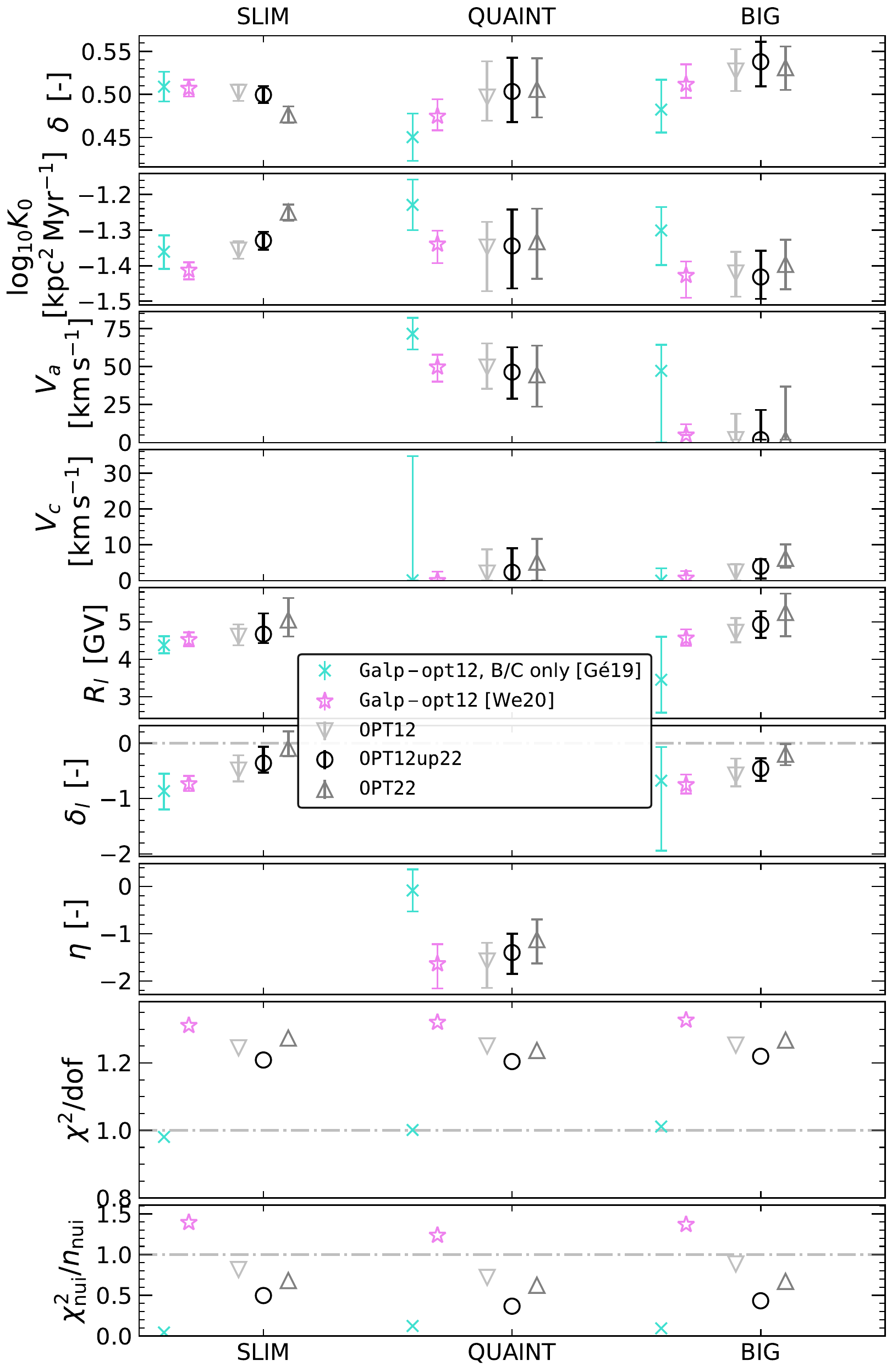}
   \caption{Same as Fig.~\ref{fig:slim_pars}, but also showing the parameter constraints for the \QUAINT{} and \BIG{} propagation configurations.
   \label{fig:allmodels_pars}}
\end{figure}

In the main text, we only discuss results for the \SLIM{} propagation configuration (see Sect.~\ref{sec:impact_transport}). For completeness, we show in Fig.~\ref{fig:allmodels_pars} the constraints set on the transport parameters in the two other configurations \QUAINT{} and \BIG{} \citep{2019PhRvD..99l3028G}; these configurations were considered in several of our previous studies \citep{2019PhRvD..99l3028G,2020PhRvR...2b3022B,2020A&A...639A..74W,2020A&A...639A.131W}, in particular in the context of dark matter analyses \citep{2021PhRvD.104h3005G,2022ScPP...12..163C}. The respective free parameters for the three configurations are: (i) for \SLIM{}, $K_0$, $\delta$, $R_l$, and $\delta_l$, that is diffusion normalisation and slope with a low-rigidity break; (ii) for \QUAINT{}, $K_0$, $\delta$, $\eta$, $V_a$, $V_c$, that is diffusion without a low-rigidity break but a possible upturn in the non-relativistic regime ($\beta$ replaced by $\beta^\eta$ in Eq.~\ref{eq:def_K}), re-acceleration, and convection; (iii) for \BIG{}, $K_0$, $\delta$, $R_l$, and $\delta_l$, $V_a$, and $V_v$, that is diffusion with a low-rigidity break, reacceleration, and convection.

As in \SLIM{}, \BIG{} and \QUAINT{} show a better \chimindof{} and \chipernui{} for all the updated cross-section sets, with \optxiiupxxii{} providing the best-fit for all transport configurations. Also, moving from \optxii{} to \optxiiupxxii{}, and \optxxii{}, the significance of the low-rigidity break decreases ($\delta_l$ consistent with zero for the latter set). However, at variance with \SLIM{}, the parameter $K_0$ is no longer sensitive to the production cross-section set considered, that is whether we use the original \xsGalxii{} set (violet stars) or the updated ones (grey and black symbols). Actually, although $V_a$ and $V_c$ are consistent for all the different cross-section sets, their best-fit values and $1\sigma$ upper limits are larger for the updated sets: convection and reacceleration dominate over diffusion below tens of GV \citep[e.g.][]{2019A&A...627A.158D,2022arXiv220306479V}, and the fit prefers to modify these parameters instead of adjusting $K_0$.
%

\bibliographystyle{aa} 
\bibliography{libeb_update}
\end{document}